\documentclass[apj]{emulateapj}

\usepackage{graphicx, longtable, latexsym, amssymb, lscape, natbib, color}

\begin{document}

\newcommand{\scprime}{{\sc prime}}
\newcommand{\Bprime}{{\sc Bprime}}
\newcommand{\Hprime}{{\sc Hprime}}
\newcommand{\HIprime}{{\sc Gasprime}}

\title{Multi-Wavelength Properties of Barred Galaxies in the Local
Universe. I: Virgo Cluster}

\author{Lea Giordano\altaffilmark{1}, Kim-Vy H. Tran\altaffilmark{1,2},
  Ben Moore\altaffilmark{1} \& Am\'{e}lie Saintonge\altaffilmark{3,4}}
\email{giordano@physik.uzh.ch}

\altaffiltext{1}{Institute for Theoretical Physics, University of
  Z\"{u}rich, Switzerland} 

\altaffiltext{2}{George P. and Cynthia W. Mitchell Institute for
Fundamental Physics and Astronomy, Department of Physics \& Astronomy,
Texas A\&M University, College Station, TX 77843}

\altaffiltext{3}{Max Planck Institute for Astrophysics, Garching,
  Germany}

\altaffiltext{4}{Max-Planck-Institut f\"{u}r Extraterrestrische
  Physik, Garching, Germany}

\begin{abstract}

  We study in detail how the barred galaxy fraction varies as a
  function of luminosity, HI gas mass, morphology and color in the
  Virgo cluster in order to provide a well defined, statistically
  robust measurement of the bar fraction in the local universe
  spanning a wide range in luminosity (factor of $\sim$100) and HI gas
  mass.  We combine multiple public data-sets (UKIDSS near-infrared
  imaging, ALFALFA HI gas masses, GOLDMine photometry).  After
  excluding highly inclined systems, we define three samples where
  galaxies are selected by their B-band luminosity, H-band luminosity,
  and HI gas mass. We visually assign bars using the high resolution
  H-band imaging from UKIDSS.  When all morphologies are included, the
  barred fraction is $\sim17-24$\% while for morphologically selected
  discs, we find that the barred fraction in Virgo is $\sim29-34$\%:
  it does not depend strongly on how the sample is defined and does
  not show variations with luminosity or HI gas mass. The barred
  fraction depends most strongly on the morphological composition of
  the sample: when the disc populations are separated into lenticulars
  ({\it S0--S0/a}), early-type spirals ({\it Sa--Sb}), and late-type
  spirals ({\it Sbc--Sm}), we find that the early-type spirals have a
  higher barred fraction ($\sim45-50$\%) compared to the lenticulars
  and late-type spirals ($\sim22-36$\%). This difference may be due to
  the higher baryon fraction of early-type discs which makes them more
  susceptible to bar instabilities. We do not find any evidence of
  barred galaxies being preferentially blue.

\end{abstract}

\keywords{galaxies:barred galaxies}

\maketitle

\section{Introduction}
\label{sec:intro}
\setcounter{footnote}{0}

Instabilities in disc galaxies play an important role as these
galaxies evolve through cosmic time. Secular and transient
instabilities can strongly affect star formation histories, disc scale
lengths, morphologies, fueling rates of central AGN etc. The bar
instability is one of the most common secular effects which is
observed in about two thirds of disc galaxies in the near infrared
\citep[][hereafter E00]{Eskridge2000}.  On scales smaller than 1 kpc,
more than half of disc galaxies host secondary bars, central star
clusters, spiral-like dust lanes, or star-forming rings
\citep{Carollo1997_P1, Carollo1998_P2, Martini1999, Laine1999,
  Boeker2003}. Bars themselves are unstable and they can buckle to
form boxy/peanut shaped bulges, similar to observed in our own Milky
Way \citep{Combes1990, Pfenniger1990, Bureau2005, Athanassoula2005,
  Martinez-Valpuesta2006, Debattista2006}.

According to recent studies \citep{VanDenBergh2002, Barazza2009}, the
bar fraction does not change within different environments but depends
solely on host-galaxy properties. \cite{Barazza2009} finds a small
increase in bar fraction only in dense central cores of clusters even
though on average the fractions are comparable: $\sim$30\% in clusters
and in the field.

In addition, there is no significant difference between the global
properties of barred and unbarred galaxies \citep{Kalloghlian1998}:
for a given circular velocity, they have comparable luminosities,
scale lengths, colors, and star formation rates
\citep{Courteau2003}. This suggests that barred and unbarred galaxies
are members of the same family and do not originate from different
evolutionary trees. Their structural similarity may be understood if
bars are generated by transient dynamical processes that are
independent of the initial galaxy formation conditions.

The main difficulty in comparing bar fractions in different
environments and at different redshifts is that, depending on the
inclination, the dust content and the color of the host galaxy, the
bar structure in the disc can be hidden at optical wavelengths.

The first statistically robust study that looked into the difference
between optical and near-infrared (NIR hereafter) bar fraction has
been made by E00.  Starting from an optically selected, magnitude
limited (B$\leq$12) sample (the OSU Bright Spiral Galaxy survey,
\citealt{Eskridge2002}), including only big (diameter
D$\geq$6$^{\prime}$) lenticular and spiral galaxies (with Hubble type
T$\geq$0) in the RC3 \citep{RC3}, they found that the NIR
bar fraction is at least double than the optical one. This leads to the
conclusion that RC3 bar types should be used with caution (see also
\citealt{Marinova2007}).

It is interesting to note that recent studies such as
\cite{Marinova2007, MenendezDelmestre2007, Barazza08, Aguerri2009,
  Marinova2009} have found, for local samples of spiral galaxies,
optical bar fractions as high as 70\% (in case of disc dominated
systems). Thus the true bar fraction could be even higher if is 
measured in NIR.

The variation in bar fraction statistics in previous studies leaves
some open questions, in particular regarding the importance of survey
wavelength and the variation as a function of sample morphology. Thus
we wish to clarify the following key issues: How does the bar fraction
vary across the late to early type disc galaxies?  Does the bar
fraction vary with redshift, i.e. do galaxies know they should become
barred once they form?  Does the bar fraction vary with environment
i.e. are bars triggered by gravitational interactions?  In order to
address the previous questions, we need to anchor the bar fraction at
redshift zero: we may be then able to shed light on why some galaxies
are barred and others are not.

Thus, the goal of this study is to provide a robust measurement of the
bar fraction at $z=0$, as a function of wavelength, Hubble type, HI
mass content and to test and quantify the impact of sample selection
on the inferred bar fraction. We quantify the NIR numbers of barred
galaxies in the Virgo cluster using newly released observations from
the UKIDSS Large Area Survey (H-band imaging). One of the main
advantages of the H-band is that it is less sensitive to dust
obscuration than B-band imaging typically used for bar studies.  We
also use HI gas masses from the ongoing ALFALFA survey to study the
gas contents of these galaxies. In Paper II of this series we will
apply the same approach to a well defined sample of field galaxies for
comparison to the cluster sample.

The outline of the present paper is as follows: we present the data in
\S \ref{subsec:data}, the galaxy classification method in \S
\ref{subsec:classification}, our sample selection in \S
\ref{subsec:sampleselection}; the results are presented in \S
\ref{sec:results} and discussed in \S \ref{sec:discussion}. We draw
conclusions in \S \ref{sec:conclusions}. Throughout the paper we
assume a flat cosmology with $\Omega_M$=0.3, $\Omega_{\lambda}$=0.7
and $H_0$=75 km s$^{-1}$ Mpc$^{-1}$. We adopt a distance modulus for
the Virgo cluster $(m-M)=30.5$ from \cite{deVaucouleurs1977} that
gives for an angular distance of 1$\arcsec$ a physical distance of 68
pc.

\section{Observations}

\label{subsec:data}

The Virgo Cluster is the nearest galaxy cluster to the Milky Way and
consists of over 2000 known members at a distance of approximately 18
Mpc \citep{Binggeli1985_P2} and a total mass of about 10$^{14}$
M$_{\odot}$ \citep{Binggeli99}.  Virgo's extensive substructure
indicates it is a dynamically young system; however, Virgo's proximity
makes it an ideal location for investigating the properties of barred
galaxies in an over-dense environment.  The main challenges in studying
galaxies in Virgo are the cluster's angular size of more than 100
deg$^2$ and that the typical angular size for individual members is
larger than the field-of-view of most CCDs.  Thus an extensive and
careful observing campaign is needed to identify and obtain accurate
photometry for Virgo galaxies.

\begin{figure}  
\centering
\includegraphics[width=0.40\textwidth]{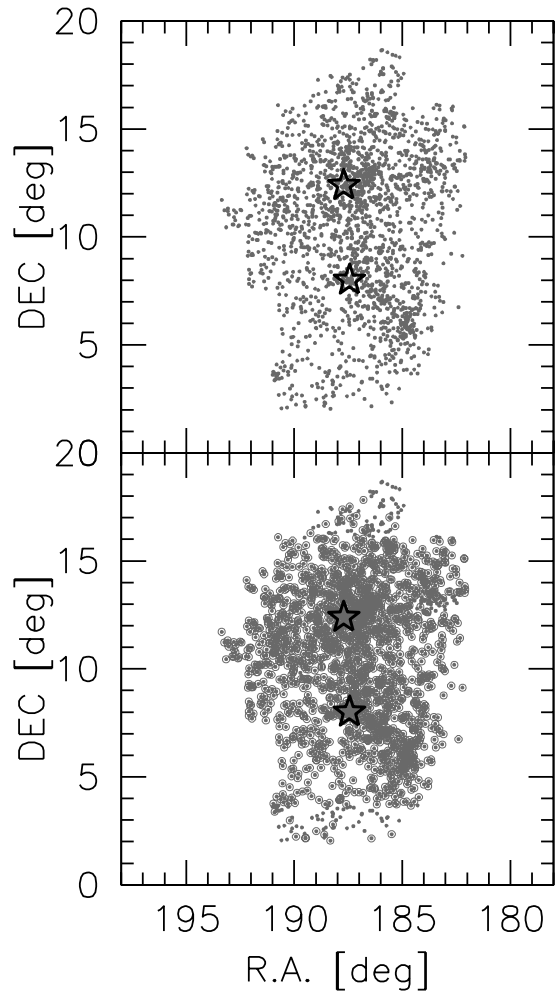}
\caption{{\it Top:} The spatial distribution of all the VCC+ galaxies
  on the sky; the two stars are M87 (top) and M49 (bottom). At the
  distance of M87 (distance modulus of $(m-M)=30.5$,
  \citealt{deVaucouleurs1977}), $1^{\circ}$ corresponds to a physical
  scale of $\sim0.24$ Mpc. {\it Bottom:} The VCC+ galaxies with
  $H$-band imaging from the UKIDSS LAS survey are shown with black
  circles. The uniform coverage of the NIR imaging confirms
  that there is no spatial bias in how our samples are selected.}
\label{fig:01}
\end{figure} 

Fortunately such a catalog already exists in the Virgo Cluster Catalog
first presented by \citet[][VCC hereafter]{Binggeli1985_P2}.  The VCC
is drawn from a long-exposure photographic plate survey covering the
range in apparent photographic blue magnitude from
$BT$\footnote{Throughout our manuscript, $BT$ magnitudes refer to the
  magnitudes measured by \citet{Binggeli1985_P2} from photographic
  plates.}$=11$ to $20$ (see \citet{Binggeli1984_P1} for details) and
the VCC has been the primary source catalog for subsequent studies of
the Virgo cluster.  We use the VCC as the master catalog throughout
our analysis.

\begin{figure*}  
\centering
\includegraphics[width=0.28\textwidth]{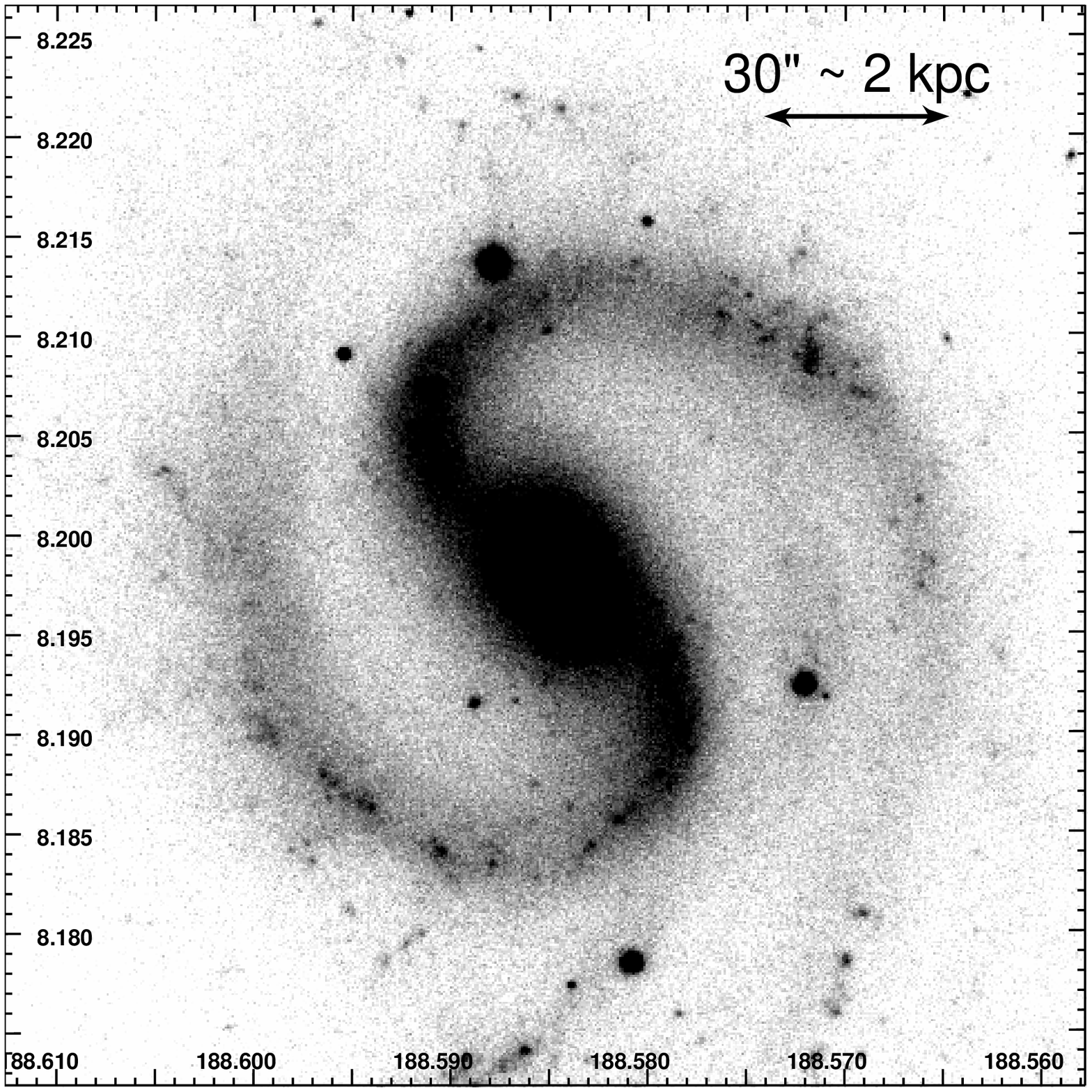}
\includegraphics[width=0.30\textwidth]{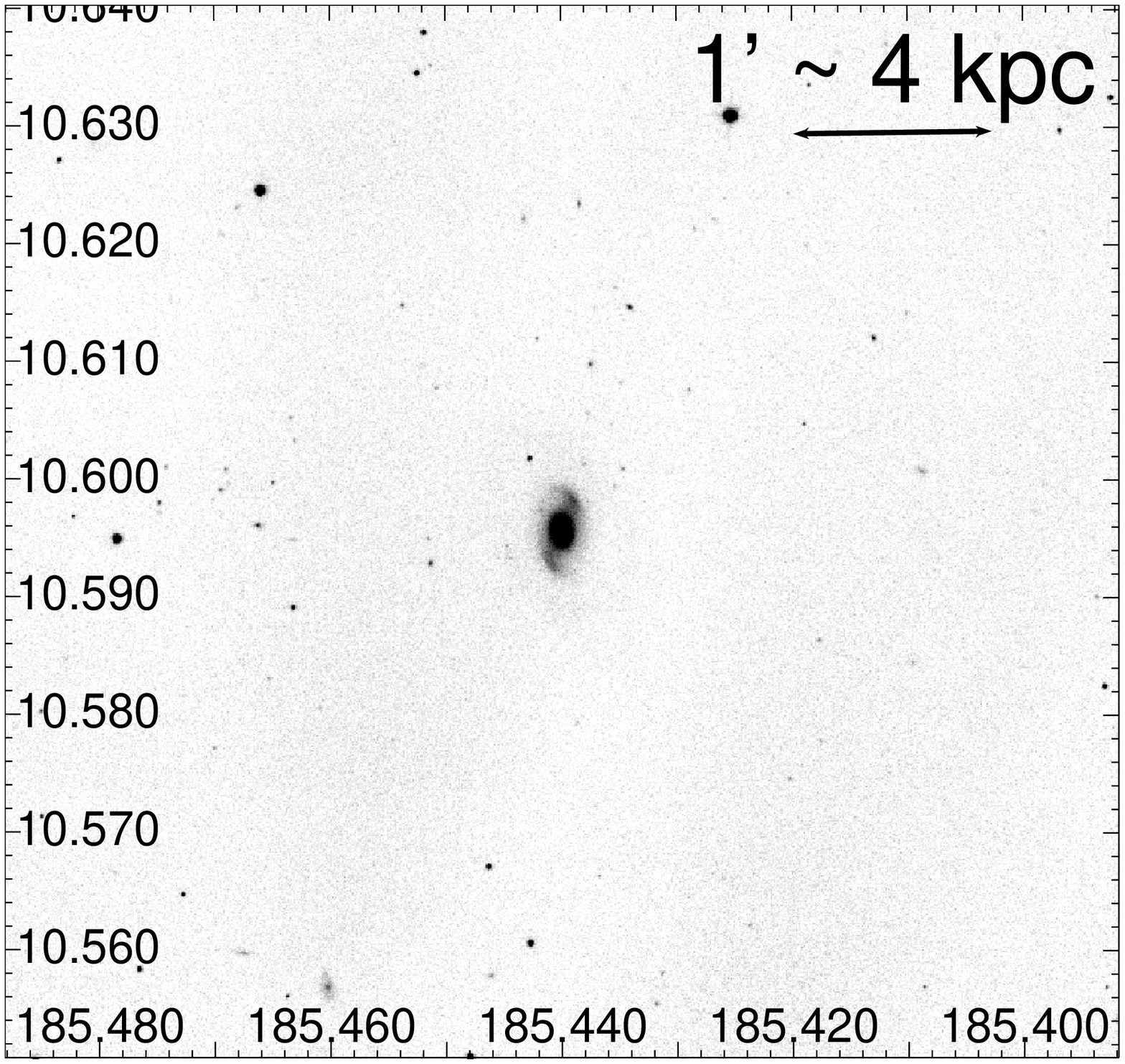}
\includegraphics[width=0.30\textwidth]{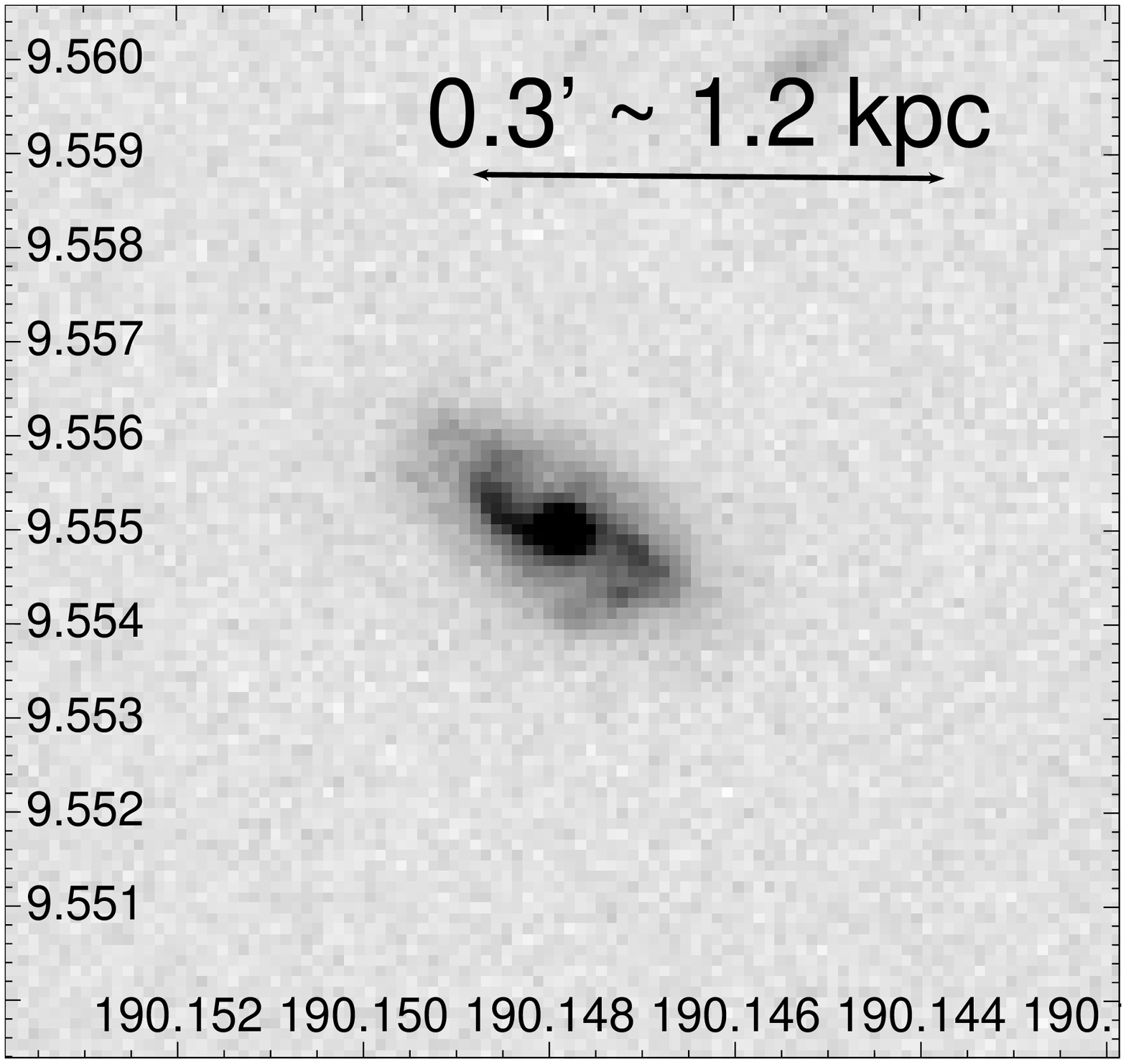}
\caption{UKIDSS LAS H-band images (0.4$^{\arcsec}$/pixel) for VCC
  1555, VCC 500 and VCC 1849, three barred galaxies in our analysis
  that have different projected sizes; the physical scale is shown as
  a black line in each panel. The high resolution of the $H$-band
  imaging enables us to identify bars across the range of luminosity
  and size spanned by Virgo members and easily includes bars smaller
  than 1~kpc.}
\label{fig:02}
\end{figure*}

While the Virgo members are drawn from the VCC, we capitalize on
recently released observations from multiple surveys in our analysis.
We focus on three different samples to investigate the bar fraction as
a function of:

\begin{enumerate}

\item B-band luminosity: Optically selected samples are affected by
  dust obscuration and are biased towards younger stellar populations.
  However, we include a B-band selected sample so that we can directly
  compare to results in the literature and thus quantify the bias
  between samples selected at different wavelengths.

\item H-band luminosity: NIR selected samples are better
  tracers of the total stellar mass in galaxies \citep{Bell01,
    Balogh01} because they are less affected by dust obscuration.
  Earlier studies (E00) indicate that the bar fraction
  measured using IR imaging is higher than in the optical.

\item HI gas mass: Selecting by total mass in neutral hydrogen has
  different selection effects, e.g. HI samples are naturally biased
  towards gas-rich galaxies that typically have spiral or irregular
  morphologies \citep{Roberts1994, Gavazzi2008}.  If bar formation is
  strongly coupled to gas mass, we expect to measure a peculiar bar
  fraction in the HI selected sample.

\end{enumerate}

\begin{figure}  
\centering
\includegraphics[width=0.48\textwidth]{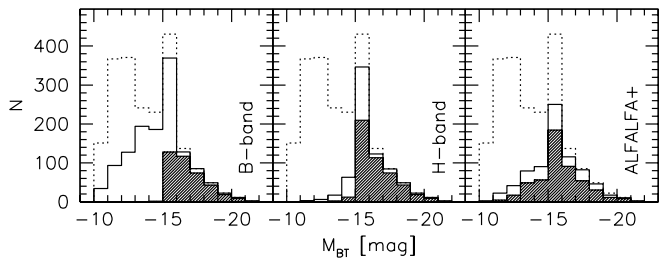}
\caption{The absolute $BT$ magnitude distribution (measured by
  \cite{Binggeli1985_P2} from photograph plates) for all the VCC+
  galaxies (dotted line histograms); note that the VCC+ is complete
  for galaxies brighter than $M_{BT}\sim-12.5$. The solid-line
  histograms show the GOLDMine $B$-band identified members (left),
  GOLDMine $H$-band identified members (middle), and ALFALFA+ HI gas
  mass identified members (right).  The shaded histograms in each
  panel are the members that have NIR imaging from UKIDSS, $BH$
  magnitudes from GOLDMine, and ALFALFA+ HI gas masses; we also have
  excluded highly inclined galaxies (axis ratio $<0.3$). The shaded
  histograms are our defined \scprime\ samples (see
  \S\ref{subsec:sampleselection}).}
\label{fig:03}
\end{figure}

\subsection{Optical Imaging}
\label{subsec:optical}

The GOLDMine database\footnote{\texttt{http://goldmine.mib.infn.it/}}
\citep{Gavazzi2003_GM} contains multi-wavelength continuum photometry,
line photometry, and dynamical and structural parameters for all 2096
Virgo galaxies in the VCC \citep{Binggeli1985_P2} as well as an
additional 60 spectroscopically confirmed members \citep{Gavazzi1999}.
We refer to the total database of 2156 Virgo galaxies as VCC+; their
spatial distribution is shown in Figure~\ref{fig:01} (top panel).

The GOLDMine observations of the VCC were constructed to be optically
complete to $BT=18.0$ ($M_{BT}=-12.5$ using a distance modulus for
Virgo of $(m-M)=30.5$).  Imaging and spectroscopic data were collected
over a 15 year period, and the ongoing observing campaign continues to
expand the multi-wavelength dataset.  From the GOLDmine database we
draw coordinates (J2000), Johnson-Cousin $B V H$ Vega magnitudes
computed within the optical radius defined by the 25 mag
arcsec$^{-2}$ isophote \citep[$\mu_{25}$;][]{Gavazzi1996}, the axis
ratios (minor to major optical diameter at $\mu_{25}$) and the
morphological type \cite[adopted from][]{Binggeli1985_P2} for the VCC+
galaxies.

\subsection{NIR Imaging}
\label{subsec:UKIDSS}

The UKIDSS Large Area Survey (LAS)\footnote{UKIDSS uses the UKIRT Wide
  Field Camera (WFCAM, \citealt{Casali2007}) and a photometric system
  described by \cite{Hewett2006}. The pipeline processing and science
  archive are described in Irwin et al. (2010, in prep.) and
  \cite{Hambly2008}.  We have used data from the 6th data release:
  \texttt{http://www.ukidss.org/surveys/surveys.html}}
\citep{Lawrence2007} is an ongoing survey to image 4000 deg$^2$ at
high Galactic latitudes in the $YJHK$ filters to a depth in $H$ of
18.8 mag; for the Virgo cluster, this limiting $H$-band magnitude
corresponds to an absolute magnitude of $M_H\sim-11.7$.  The UKIDSS
spatial resolution of 0.4$\arcsec$, average seeing of 0.8$\arcsec$,
and coverage of the Virgo cluster is unprecedented.  In comparison,
the all-sky 2MASS survey \citep{2MASS} imaged the Virgo cluster with a
pixel scale of 2$\arcsec$, has a magnitude limit of $m_H=14.3$, and is
unable to resolve features smaller than $\sim$140 pc.

Galaxy bars in the local universe tend to be smaller than 1 kpc
\citep{Laine2002, Barazza08}; this corresponds to an angular size of
$\sim15''$ at Virgo's distance, a size that is well-sampled by the
UKIDSS imaging.  Virgo's proximity combined with the UKIDSS resolution
means that we can easily resolve bars down to sub-kpc
scales. Figure~\ref{fig:02} shows UKIDSS $H$-band imaging for
three Virgo galaxies spanning the range in size and includes the
smallest barred galaxy in Virgo (diameter$\sim20''$).

We note that while we use the UKIDSS imaging to identify galaxy bars,
we use the $H$-band magnitudes from GOLDmine for the photometry.  The
existing UKIDSS data reduction pipeline does not measure the
background correctly for very extended sources such as Virgo galaxies,
thus the UKIDSS photometry for these objects is known to be incorrect
(Richard McMahon, private communication).

We queried the UKIDSS LAS database for the VCC+ galaxies and retrieved
the available H-band stacked frames (15$\arcmin$ $\times$
90$\arcmin$).  In several cases, the Virgo galaxies are on multiple
frames and the frames are stitched together using \texttt{swarp}
\citep{Bertin2002}.  Figure~\ref{fig:01} (bottom panel) shows the
spatial distribution of the VCC+ galaxies for which we also have
$H$-band imaging from the UKIDSS LAS.  The uniform coverage of the
$H$-band imaging compared to the optical indicates that there is no
spatial bias in how our samples are selected (see
\S\ref{subsec:sampleselection}).

\begin{figure}  
\centering
\includegraphics[width=0.48\textwidth]{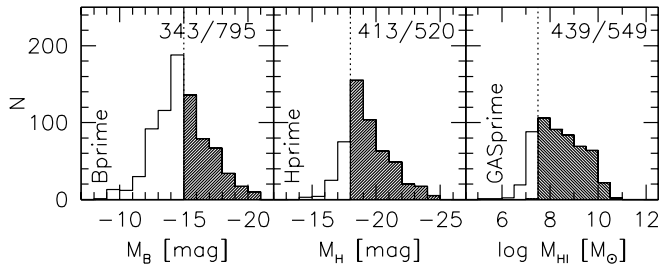}
\caption{We show the GOLDMine $B$-band magnitude distribution for our
  \Bprime\ sample (left), the GOLDMine $H$-band magnitude distribution
  for our \Hprime\ sample (middle), and the ALFALFA+ HI gas mass
  distribution for our \HIprime\ sample (right) as shaded histograms;
  for explanation of the open histograms, see
  Figure~\ref{fig:03} and \S\ref{subsec:sampleselection}.  The
  dotted vertical lines show the limiting cut for the \scprime\
  samples, i.e. $M_B\leq-15$ mag (left), $M_H\leq-18$ mag (middle),
  and $M_{HI}\geq10^{7.5}M_{\odot}$ (right). The total number of
  galaxies in each galaxy sample (open histogram) and the number in
  the \scprime\ sample (shaded histogram) are included in each panel.}
\label{fig:04}
\end{figure} 

\begin{deluxetable*}{lrrrr}      
\tabletypesize{\scriptsize}
\tablewidth{0.80\textwidth}
\tablecaption{Morphological Composition of the \scprime\ Samples\tablenotemark{a}}
\centering
\tablehead{
\colhead{ }  & \colhead{VCC+} & \colhead{\Bprime } &
  \colhead{\Hprime } & \colhead{\HIprime }
} 
\startdata
    all galaxies            & 2156          & 343         & 413          & 439          \\
    \hline
    \noalign{\smallskip}
    Dwarfs ({\it dE, dS0, dE/S0})     & 1175 ($\sim$55\%) & 46 ($\sim$13\%) & 58  ($\sim$14\%) & 35   ($\sim$8\%) \\
    Ellipticals ({\it E--E/S0})       & 67    ($\sim$3\%) & 38 ($\sim$11\%) & 44  ($\sim$11\%) & 13   ($\sim$3\%) \\
    Lenticulars ({\it S0--S0/a})      & 136   ($\sim$6\%) & 69 ($\sim$20\%) & 94  ($\sim$23\%) & 33   ($\sim$8\%) \\
    Early-type Spirals ({\it Sa--Sb}) & 128   ($\sim$6\%) & 53 ($\sim$16\%) & 77  ($\sim$18\%) & 53  ($\sim$12\%) \\
    Late-type Spirals ({\it Sbc--Sm}) & 236  ($\sim$11\%) &109 ($\sim$32\%) & 111 ($\sim$27\%) & 150 ($\sim$35\%) \\
    Irregulars/Peculiars        & 414  ($\sim$19\%) & 28  ($\sim$8\%) & 29   ($\sim$7\%) & 150 ($\sim$34\%) \\
\enddata
\tablenotetext{a}{~Listed is the total number for a given morphological
  class and its relative fraction to the total number of galaxies (in
  parentheses).}
\label{tab:01}
\end{deluxetable*}

\subsection{HI Gas Masses}
\label{subsec:ALFALFA}

The Arecibo Legacy Fast ALFA
(ALFALFA\footnote{\texttt{http://egg.astro.cornell.edu}})
\citep{Giovanelli2005_ALFALFA} survey is an on-going blind
extra-galactic search for neutral hydrogen (HI) that will cover 7074
degrees$^2$ of the high galactic latitude sky accessible to the
Arecibo telescope, i.e. $03h<\alpha<22h$ and
$0^{\circ}<\delta<+36^{\circ}$.  The ALFALFA survey covers the entire
Virgo Cluster with a spatial accuracy of $\sim20\arcsec$ for the
centroids of galaxies, a beam-size of 3.5$\arcmin$, and an HI-mass
detection limit of $M_{HI}\geq10^{7.5}M_{\odot}$.  For our analysis,
we obtain HI masses for all the available VCC+ galaxies from the
current ALFALFA data release which covers the Virgo region at
$\delta>+8^{\circ}$ \citep{Giovanelli2007, Kent2009}.

The GOLDmine database also contains HI data taken with Arecibo, the
{\it Very Large Array}, and the Nan\c{c}ay radio telescope for the
VCC+.  We integrate these HI observations \citep{Gavazzi2005} with
those from the ALFALFA survey to obtain HI masses for Virgo galaxies
at $\delta < +8^{\circ}$; we refer to the combined HI catalog as
ALFALFA+.

\begin{deluxetable*}{lrrrrr}    
\tabletypesize{\scriptsize}
\tablewidth{0.80\textwidth}
\tablecaption{Barred Galaxy Fractions\tablenotemark{a}}
\centering
\tablehead{
\colhead{Sample} & \colhead{$N$} & \colhead{$N_{Bar}$} & \colhead{Barred \%}  & \colhead{$N_{Unc.}$} & \colhead{Unc. \%}  \\ 
}
\startdata
\Bprime\  -- All\tablenotemark{b} & 343 & 78 & 22.7$\pm$2.8\% & 28 & 8.2$\pm$1.6 \% \\
\Hprime\  -- All\tablenotemark{b} & 413 &100 & 24.2$\pm$2.7\% & 40 & 9.7$\pm$1.6 \%\\
\HIprime\ -- All\tablenotemark{b} & 439 & 75 & 17.1$\pm$2.1\% & 44 &10.0$\pm$1.6 \% \\
\hline
\Bprime\  -- Discs Only\tablenotemark{c} & 231 & 77 & 33.3$\pm$4.4\% & 24 & 10.4$\pm$2.2\% \\
\Hprime\  -- Discs Only\tablenotemark{c} & 282 & 97 & 34.4$\pm$4.0\% & 37 &  9.6$\pm$2.3\% \\
\HIprime\ -- Discs Only\tablenotemark{c} & 241 & 70 & 29.0$\pm$3.9\% & 36 & 14.9$\pm$2.7\% \\
\enddata
\tablenotetext{a}{All bar classifications assigned using UKIDSS
    H-band imaging.  The ``uncertain'' class contains members for
    which there was no unanimous agreement on the classification;
    these make up $<10$\% in each of the \scprime\ samples.  In our
    analysis, the uncertain galaxies are included with the non-barred
    galaxies.}
  \tablenotetext{b}{The bar fraction relative to all galaxies in the
    \scprime\ sample.}
  \tablenotetext{c}{The bar fraction relative to only the \scprime\ disc
    galaxies  ({\it S0--Sm}).}
\label{tab:02}
\end{deluxetable*}

\begin{figure}  
\centering
\includegraphics[width=0.40\textwidth]{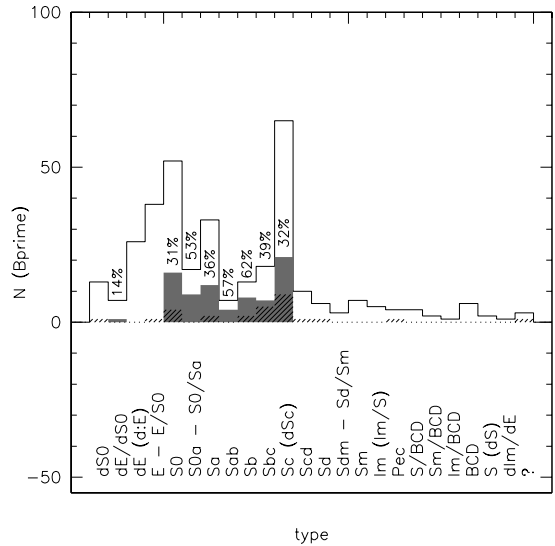}
\caption{The morphological distribution of Virgo members in our
\Bprime\ sample (solid line).  The filled histogram represents the
galaxies classified as barred using UKIDSS H-band imaging; the barred
fraction in each bin is quoted.  The hatched histogram represents the
galaxies with ``uncertain'' bar classification; in our analysis, these
are grouped with the non-barred members.}
\label{fig:05}
\end{figure} 

\begin{figure}  
\centering
\includegraphics[width=0.40\textwidth]{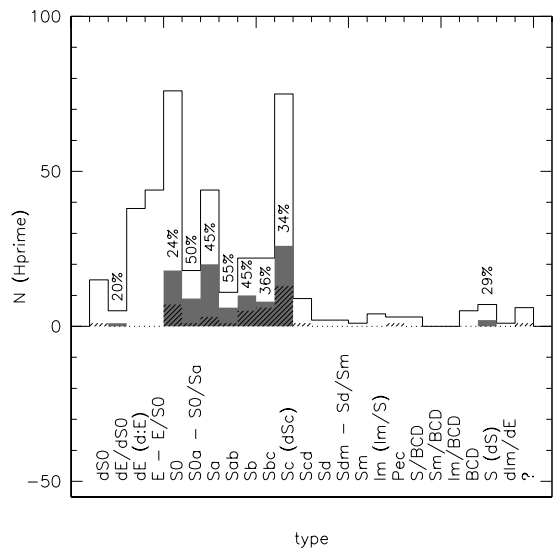}
\caption{Same as in Figure~\ref{fig:05} except for our \Hprime\
sample.  Despite selecting in the infrared instead of the optical, the
\Hprime\ sample has essentially the same morphological distribution as
the \Bprime\ sample.}
\label{fig:06}
\end{figure} 

\begin{figure}  
\centering
\includegraphics[width=0.40\textwidth]{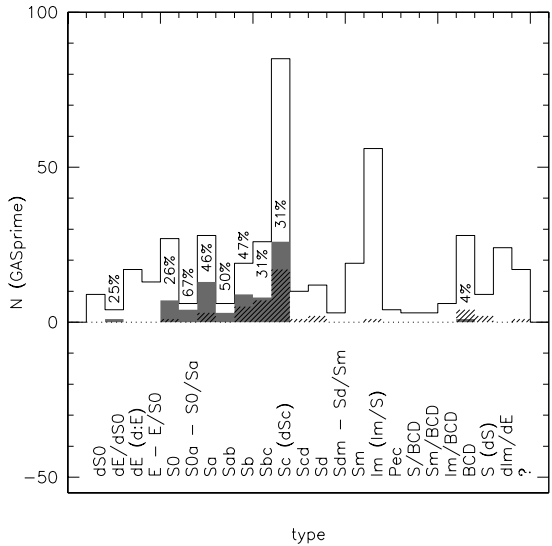}
\caption{Same as in Figure~\ref{fig:05} except for our \HIprime\
sample.  The morphological distribution of the \HIprime\ sample is very
different from the \Bprime\ and \Hprime\ samples: selecting by HI
gas mass includes many more late-type spiral and irregular galaxies
and significantly fewer elliptical and lenticular galaxies.}
\label{fig:07}
\end{figure}

\begin{deluxetable*}{lllllllll}  
\tabletypesize{\scriptsize} 
\tablewidth{0.80\textwidth}
\tablecaption{Virgo galaxy catalogue for \scprime\ samples\tablenotemark{a}.}  
\tablehead{ 
  \colhead{Galaxy} & \colhead{\Bprime } & \colhead{\Hprime} &
  \colhead{\HIprime} & \colhead{M$_B$\tablenotemark{b}} &
  \colhead{M$_H$}\tablenotemark{c} &
  \colhead{M$_{HI}$\tablenotemark{d}} & \colhead{type
    \tablenotemark{e}} & \colhead{class \tablenotemark{f}} 
}
\startdata
  VCC    3 & no & no & yes&  -13.50  &  \nodata  & 8.67    & IRR & no      \\
  VCC    4 & no & no & yes&  -13.13  &  \nodata  & 8.25    & IRR & no      \\
  VCC    6 & no & yes& no &  -14.81  &  -18.30   & \nodata & ETS & yes     \\
  VCC    9 & yes& yes& no &  -16.23  &  -19.06   & 7.45    & DWF & no      \\
  VCC   12 & no & yes& yes&  -14.76  &  -18.64   & 8.04    & ETS & yes     \\
  VCC   14 & no & no & yes&  -13.52  &  \nodata  & 9.41    & IRR & no      \\
  VCC   16 & no & no & yes&  \nodata &  \nodata  & 9.08    & LTS & no      \\
  VCC   17 & no & no & yes&  -14.22  &  -16.20   & 8.78    & IRR & no      \\
  VCC   18 & yes& yes& yes&  -15.20  &  -18.93   & 9.99    & LTS & no      \\
\enddata
\tablenotetext{a}{Rest of table available as online material.}
\tablenotetext{b}{B-band magnitude from GOLDMine.}
\tablenotetext{c}{H-band magnitude from GOLDMine.}
\tablenotetext{d}{HI gas mass (in units of log M$_{\odot}$) from
  ALFALFA+.}  \tablenotetext{e}{Morphological classification from
  VCC+: DWF ({\it dE, dS0, dE/S0}), ELL ({\it E-E/S0}), LEN ({\it S0-S0a}), ETS ({\it Sa-Sb}),
  LTS ({\it Sbc-Sm}), IRR ({\it Im, Pec, S/BCD, Sm/BCD, Im/BCD, BCD, dS, dIm,
  ?}).}  
\tablenotetext{f}{Bar classifications assigned using UKIDSS H-band imaging.}
\label{tab:03}
\end{deluxetable*}

\subsection{Defining the \scprime\ Samples}
\label{subsec:sampleselection}

One of our goals is to determine if the galaxy bar fraction depends on
how the galaxy samples are defined. In our analysis, we consider only
galaxies that are covered by the UKIDSS imaging with measured $B$
(optical) and $H$ (NIR magnitudes from GOLDMine) and HI gas
masses.  Figure~\ref{fig:03} shows the absolute magnitude
($M_{BT}$) distribution of the VCC+ galaxies (dotted histogram).  The
solid histograms in the three panels of Figure \ref{fig:03}
show the 1) GOLDMine $B$-magnitude identified members (left), 2)
GOLDMine $H$-magnitude identified members (middle), and 3) HI gas mass
identified members (right). The shaded histograms in each panel
represent the galaxies used in our analysis; note that our galaxy
samples are well-above the VCC+ completeness limit of $M_{BT} \sim
-12.5$ \citep{Binggeli1985_P2}.

In our analysis, we exclude highly inclined galaxies (those with
axis ratio smaller than 0.3) because edge-on bars are difficult to
identify.  Lastly, to make up our \scprime\ samples we use only
members brighter than the VCC+ completeness limit of $M_{BT}\sim-12.5$
and we also apply a luminosity or HI gas mass cut to minimize any
potential bias due to incompleteness.  Our \scprime\ samples are:

\begin{itemize}

\item \Bprime: Virgo galaxies are selected by their GOLDMine $B$-band
  magnitude and we consider only galaxies brighter than $M_B=-15$ mag
  (343 members); Figure~\ref{fig:04} (left) shows the GOLDMine
  $B$-band magnitude distribution of this sample.  We compare results
  with the \Bprime\ sample to results from recent {\it
    optically-selected} studies.

\item \Hprime: Virgo galaxies are selected by their GOLDMine $H$-band
  magnitude and we consider only galaxies brighter than $M_H=-18$ mag
  (413 members); Figure~\ref{fig:04} (middle) shows the GOLDMine
  $H$-band magnitude distribution of this sample. Selecting by
  $H$-band luminosity should provide a more robust determination of
  the bar fraction because the $H$-band is less affected by dust.  The
  \Bprime\ and \Hprime\ samples overlap significantly; most of the
  \Hprime\ members that are not in the \Bprime\ sample are spirals
  with high dust content.

\item \HIprime: Virgo galaxies are selected by their ALFALFA+ HI gas
  mass (neutral hydrogen) and we consider only galaxies with
  M$_{HI}\geq10^{7.5}$~M$_{\odot}$ (439 members);
  Figure~\ref{fig:04} (right) shows the HI gas mass distribution
  of this sample.  If HI gas mass is correlated with bar formation, we
  expect the bar fraction in the \HIprime\ sample to be highest.

\end{itemize}

\begin{deluxetable*}{lcccccc}   
\tabletypesize{\scriptsize}
\tablewidth{0.80\textwidth}
\tablecaption{Barred Fraction\tablenotemark{a} as a Function of Luminosity \& HI Mass}
\tablehead{
\multicolumn{1}{c}{}&
\multicolumn{2}{c}{\Bprime\tablenotemark{b}} & 
\multicolumn{2}{c}{\Hprime\tablenotemark{c}} & 
\multicolumn{2}{c}{\HIprime\tablenotemark{d}}
}
\startdata
All Galaxies & $M_B$ & Barred \%& $M_H$ & Barred \% & $M_{HI}$ & Barred \%   \\
\hline
\noalign{\smallskip}
&-15 mag & 22.7\% & -18 mag & 24.3\% & 10$^{7.5}$ M$_{\odot}$ & 17.1\%    \\
&-16 mag & 29.0\% & -19 mag & 27.5\% & 10$^{8.5}$ M$_{\odot}$ & 25.1\%    \\
&-17 mag & 34.4\% & -20 mag & 33.1\% & 10$^{9.5}$ M$_{\odot}$ & 26.1\%    \\
\hline
\hline
\noalign{\smallskip}
Discs Only & $M_B$ & Barred \% & $M_H$ & Barred \% & $M_{HI}$ & Barred \%   \\
\hline
\noalign{\smallskip}
&-15 mag & 34.8\% & -18 mag & 34.8\% & 10$^{7.5}$ M$_{\odot}$ & 32.0\%   \\
&-16 mag & 36.6\% & -19 mag & 36.8\% & 10$^{8.5}$ M$_{\odot}$ & 34.7\%   \\
&-17 mag & 39.6\% & -20 mag & 40.5\% & 10$^{9.5}$ M$_{\odot}$ & 33.8\%   \\
\enddata
\tablenotetext{a}{All bar classifications assigned using UKIDSS H-band
  imaging.} 
\tablenotetext{b}{Barred Fraction when considering only \Bprime\
  galaxies brighter than this magnitude limit.}
\tablenotetext{c}{Barred Fraction when considering only \Hprime\
  galaxies brighter than this magnitude limit.}
\tablenotetext{d}{Barred Fraction when considering only \HIprime\
  galaxies with HI masses greater than this limit.}
\label{tab:04}
\end{deluxetable*}

\begin{deluxetable*}{lrrrrrr}   
\tabletypesize{\scriptsize}
\tablewidth{0.80\textwidth}
\tablecaption{Barred Galaxy Fraction: Red Sequence vs. Blue Cloud\tablenotemark{a}}
\tablehead{
\colhead{Sample}& \colhead{$N$(RS\tablenotemark{b})} & \colhead{$N_{Bar}$(RS\tablenotemark{b})}
  & \colhead{Barred \% (RS\tablenotemark{b})}  &
  \colhead{$N$(BC\tablenotemark{c})} &
  \colhead{$N_{Bar}$(BC\tablenotemark{c})} & \colhead{Barred \%
    (BC\tablenotemark{c})}
}
\startdata
  \Bprime\  -- All\tablenotemark{d} & 230 & 59 & 25.7$\pm$3.7\% &  91 & 19 & 20.9$\pm$5.3\% \\
  \Hprime\  -- All\tablenotemark{d} & 298 & 74 & 24.8$\pm$3.2\% &  84 & 19 & 22.6$\pm$5.6\% \\
  \HIprime\ -- All\tablenotemark{d} & 161 & 46 & 28.6$\pm$4.8\% & 108 & 20 & 18.5$\pm$4.5\% \\
\enddata
  \tablenotetext{a}{All bar classifications assigned using UKIDSS H-band
    imaging.} 
  \tablenotetext{b}{Galaxies on the Red Sequence (RS) as defined by the
    CM relation (\S\ref{sec:cmd}).}
  \tablenotetext{c}{Galaxies in the Blue Cloud (BC) as defined by the CM
    relation (\S\ref{sec:cmd}).}
  \tablenotetext{d}{The bar statistics in the Red Sequence or Blue
    Cloud using all morphological types in the \scprime\ sample.}
\label{tab:05}
\end{deluxetable*}

\begin{figure}  
\centering
\includegraphics[width=0.48\textwidth]{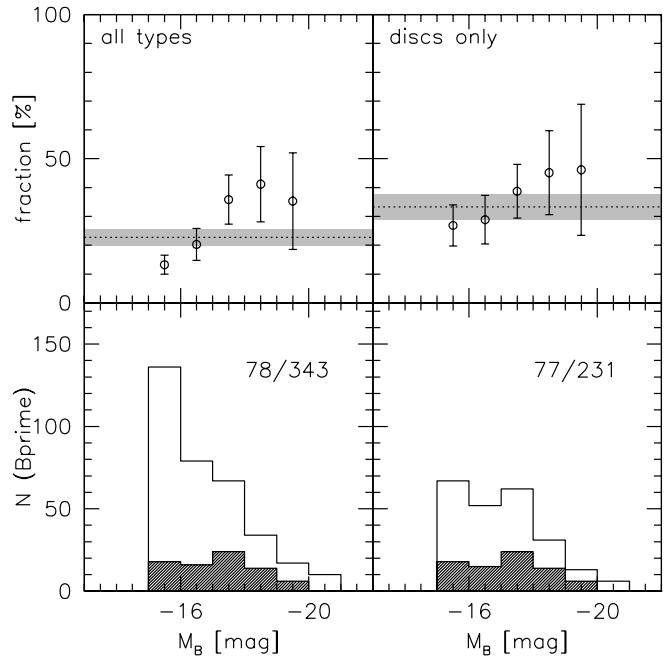} 
\caption{{\it Top panels:} The fraction of barred galaxies in the
  \Bprime\ sample as a function of luminosity for all morphological
  types (left) and only disc galaxies ({\it S0--Sm}; right); morphological
  classifications are from the VCC+. Note that our galaxy sample spans
  a factor of $\sim100$ in luminosity. The dotted line shows the
  average barred fraction in the \Bprime\ sample and the gray shaded
  region the statistical error. The error bars are the statistical
  Poisson error per bin. {\it Bottom panels:} Corresponding histograms
  of the luminosity distribution for all morphological types (left)
  and discs only (right); the shaded histograms show the barred
  members. Included in these panels are the number of barred and total
  galaxies in the sample. There is a break in the barred fraction for
  the total \Bprime\ sample at $M_B\sim-17$ mag where the barred
  fraction is lower for fainter galaxies; however, the break is weaker
  in the \Bprime\ disc only sample. We find that the marginally
  detected trend of
  decreasing barred fraction with decreasing luminosity is due
  to the numerous population of faint dwarf galaxies ($M_B \gtrsim
  -16$ mag) of which $<1$\% are barred (see also Figure
  \ref{fig:05}).}
\label{fig:08} 
\end{figure}

\begin{figure}  
\centering
\includegraphics[width=0.48\textwidth]{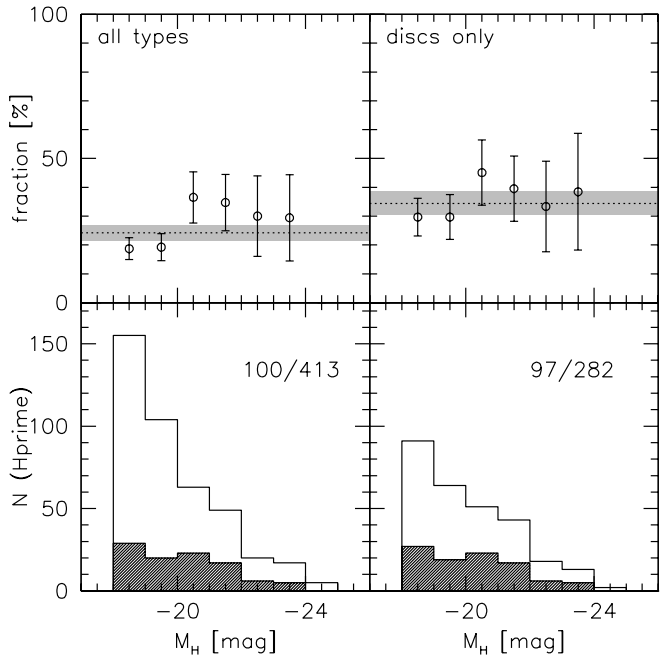} 
\caption{Same as Figure \ref{fig:08} but for the \Hprime\ sample.
  As in the \Bprime\ sample, the barred fraction is lower in the two
  faintest luminosity bins ($M_H>-20$) but fairly constant at higher
  luminosities.  Because the change in barred fraction with luminosity
  is more evident in the total \Hprime\ sample, it supports our
  conclusion that the lower barred fraction at fainter luminosities is
  due to the numerous population of faint dwarf galaxies.}
\label{fig:09}
\end{figure}

\begin{figure}  
\centering
\includegraphics[width=0.48\textwidth]{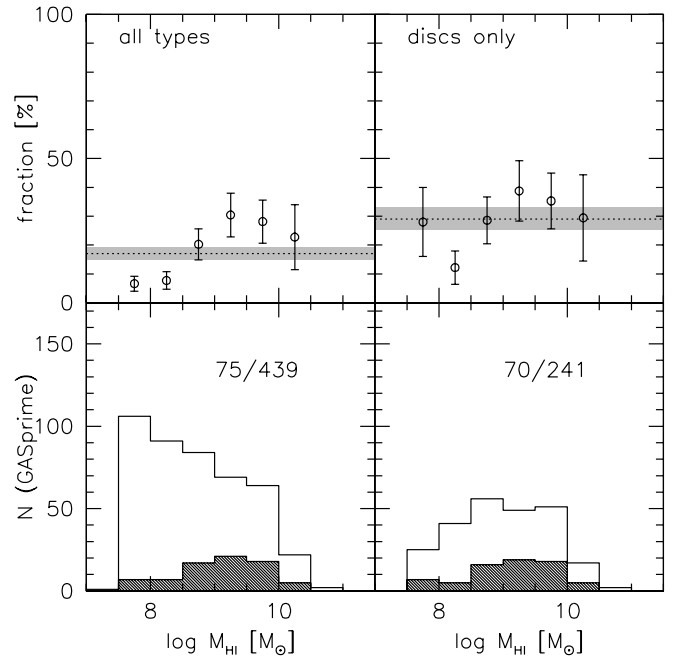} 
\caption{Same as Figs. \ref{fig:08} \& \ref{fig:09} but for the
  \HIprime\ sample.  Similar to results in the \Bprime\ and \Hprime\
  samples, the barred fraction is lowest for the galaxies with the
  smallest HI gas masses ($<10^{8.5}M_{\odot}$) and fairly constant
  for galaxies with more HI.  The lower barred fraction at lower HI
  gas mass is  due to the numerous population of dwarf galaxies,
  in particular the irregular galaxies in the \HIprime\ sample.}
\label{fig:10}
\end{figure}

\subsection{Classification Method}
\label{subsec:classification}

As part of our analysis, we use the detailed morphological types from
the VCC+.  For simplicity, we define six broad morphological classes:
dwarfs ({\it dE, dS0, dE/S0}), ellipticals ({\it E, E/S0}), lenticulars ({\it S0, S0a};
defined as red, featureless discs with very little to no sign of star
formation), early spirals ({\it Sa, Sab, Sb}; defined as spiral galaxies
with a dominant bulge component), late spirals ({\it Sbc, Sc, Scd, Sd, Sm};
defined as spiral galaxies with small bulges or bulge-less), and
irregulars/peculiars/blue compact dwarfs ({\it Im, Pec, S/BCD, Sm/BCD,
Im/BCD, BCD, dS, dIm, ?}).  The morphological make-up of the three
\scprime\ samples is listed in Table~\ref{tab:01}.

Recognizing that there are multiple approaches on how to identify
galaxy bars, we stress that the strength of our analysis lies in our
comparisons being internally consistent.  We adopt the
straight-forward method of visually classifying galaxies into one of
three categories: ``barred'', ``non-barred'', or ``uncertain''.  Note
that the ``barred'' category includes both strong, elongated bars (SB)
to weak oval bars (SAB).  Because earlier work by E00 found that the
frequency of barred galaxies in the near infrared is nearly twice that
measured in the optical, we use the $H$-band images from UKIDSS to
visually identify galaxy bars for all of our \scprime\ samples.

Three of us (LG, KT, \& BM) identified bars by inspecting the UKIDSS
$H$-band imaging for the three \scprime\ samples; LG and BM repeated
the procedure twice with a general agreement of $\sim92$\%.  If there
was unanimous agreement, we adopted the classification.  If not, the
classification was ``uncertain''; note that fewer than 10\% of
\scprime\ galaxies have ``uncertain'' bar classifications (see Table
\ref{tab:02}).  In the following analysis, we group the
``uncertain'' galaxies with the non-barred ones.
Table~\ref{tab:03} lists the photometry, HI gas mass,
morphological type, and bar classification for the Virgo galaxies in
our \scprime\ samples.

Figures~\ref{fig:05}, \ref{fig:06}, \& \ref{fig:07} show
the morphological distributions of the \Bprime, \Hprime, and \HIprime\
samples respectively, and we include the barred, non-barred, and
uncertain classifications as assigned by us.  The Virgo cluster does
have a number of elliptical galaxies (Table~\ref{tab:01}); however,
the majority of members are lenticular and spiral galaxies.  In our
analysis, we group S0 to Sm galaxies together as disc members (e.g. in
Table~\ref{tab:02}).  Note that the morphological distribution of
the \Bprime\ and \Hprime\ samples are very similar, but that the
\HIprime\ sample includes many more late-type spiral and irregular
galaxies and significantly fewer elliptical and lenticular galaxies.

\section{Results}
\label{sec:results}

In this section we present how the barred galaxy fraction depends on
selection method in our \scprime\ samples.  We compare barred
fractions for all and disc-only populations as defined by morphology
and color.  In our analysis, we calculate Poissonian statistical
errors following \cite{Gehrels86}.

\subsection{Barred Fraction: All Galaxies vs. Discs Only}

If we consider all galaxies in the \scprime\ samples, we find that the
barred fraction ranges from 17\% (\HIprime, Table~\ref{tab:02})
to 24\% (\Hprime).  The similar barred fractions are surprising given
the different morphological compositions of the \scprime\ samples:
only $\sim50$\% of galaxies in the \HIprime\ sample are classified as
discs ({\it S0--Sm}) compared to $\sim$67\% in both the \Bprime\ and
\Hprime\ samples (see Table~\ref{tab:01}).  Note that regardless of
selection method, virtually all of the barred galaxies have
morphological type {\it S0--Sm}.

The barred fraction is traditionally defined as the number of barred
discs over the total number of disc galaxies, and the most common
method for excluding non-disc galaxies is to use morphological
information.  We consider only the morphologically-selected disc
galaxies defined as having Hubble type S0 to Sm, i.e. galaxies with a
clearly defined disc component.  The barred fraction for disc galaxies
is lowest in the \HIprime\ sample ($\sim29$\%) and highest in the
\Hprime\ sample ($\sim34$\%).  However, the barred fraction does not
vary strongly across the \scprime\ samples even when considering only
disc galaxies (Table~\ref{tab:02}).

We find that the barred galaxy fraction in Virgo remains remarkably
similar across the three \scprime\ samples both when considering only
disc galaxies and when including all morphological types
(Table~\ref{tab:02}): the traditionally defined barred fraction
(discs only) is $\sim30$\% while the total barred fraction (all
morphological types) is $\sim20$\%.  We find that the barred fraction
for discs in the Virgo cluster is only about half that measured by E00
in their heterogeneous sample of discs.

\subsection{Barred Fraction vs. Luminosity}\label{sec:luminosity}

The galaxies in our \scprime\ samples span a wide range in luminosity
(factor of $\sim100$), thus we can test whether the barred galaxy
fraction varies with luminosity.  In the \Bprime\ sample, there is a
break at $M_B\sim-17$ such that the barred fraction is lower for
fainter systems (Figure \ref{fig:08}); the break is most evident when
considering all morphological types. If we apply magnitude cuts to the
\Bprime\ sample (Table~\ref{tab:04}), we find that the barred
fraction (all galaxies) increases from 23\% ($M_B<-15$ mag) to 34\%
($M_B<-17$ mag).  At brighter luminosities, the barred fraction
remains constant.  The break is weaker when considering only \Bprime\
discs (increases from 35\% to 40\%; Table~\ref{tab:04}).

We note that dwarf and irregular galaxies can have bar signatures
\citep{Barazza02,Lisker07} due to, e.g. morphological transformation
of discs into dwarf ellipticals and spheroids by the cluster potential
\citep{Moore1996Nature, Mastropietro05}.  However, the barred fraction
for these low-mass galaxies is significantly lower: \cite{Lisker07}
find that $<10$\% (37/413) have signatures of bars.  Among the
galaxies in our \scprime\ samples, we do have a number of dwarf and
irregular galaxies (see Figures~\ref{fig:05}, \ref{fig:06},
\ref{fig:07}).

In the \Bprime\ sample, we find that the lower barred fraction at
fainter luminosities is due to the population of dwarf galaxies; these
are excluded when considering only discs which explains the weaker
break observed in the disc-only sample (Figure~\ref{fig:08};
Table~\ref{tab:04}).  The dwarf galaxy population is numerous (it
makes $\sim 13$\% of the \Bprime\ sample) but has a very low barred
fraction: only two of the dwarf galaxies have bars (see
Figure~\ref{fig:05}).

We find the same results in the \Hprime\ sample for both the total and
disc-only samples (Figure~\ref{fig:09}; Table~\ref{tab:04}): there
is a break at $M_H\sim-20$ where the barred fraction at fainter
luminosities is lower, and the barred fraction is higher and remains
fairly constant at brighter luminosities.  As noted earlier, most of
the \Hprime\ sample overlaps with the \Bprime\ sample, thus the lower
barred fraction at lower luminosities is due to a similar dwarf galaxy
population (see Figure~\ref{fig:06}).

The barred fraction in the \HIprime\ sample also decreases
significantly at low HI gas masses (Figure~\ref{fig:10}).  This is
surprising given the different morphological make-up of the \HIprime\
sample compared to the luminosity-selected \Bprime\ and \Hprime\
samples (compare Figure~\ref{fig:07} to Figures~\ref{fig:05} \&
\ref{fig:06}): the \HIprime\ sample has a measurably higher
fraction of irregular galaxies and lower fraction of lenticulars (see
Table~\ref{tab:01}).  However, when considering only disc galaxies,
the barred fraction in the \HIprime\ sample behaves as in the \Bprime\
and \Hprime\ samples, i.e. it does not vary significantly with
luminosity.

In summary, all of the \scprime\ samples show a decrease in the barred
fraction with decreasing luminosity or HI gas mass because the barred
fraction in non-disc, low-mass galaxies is lower.  In the case of the
luminosity-selected samples (\Bprime\ \& \Hprime), the lower barred
fraction is due to the numerous dwarf population while in the HI
gas-mass sample (\HIprime), it is due to the irregulars.  If only disc
galaxies are considered, the barred fraction is relatively constant
over the luminosity and HI gas mass ranges of our \scprime\ samples.

\begin{figure} 
\centering
\includegraphics[width=0.40\textwidth]{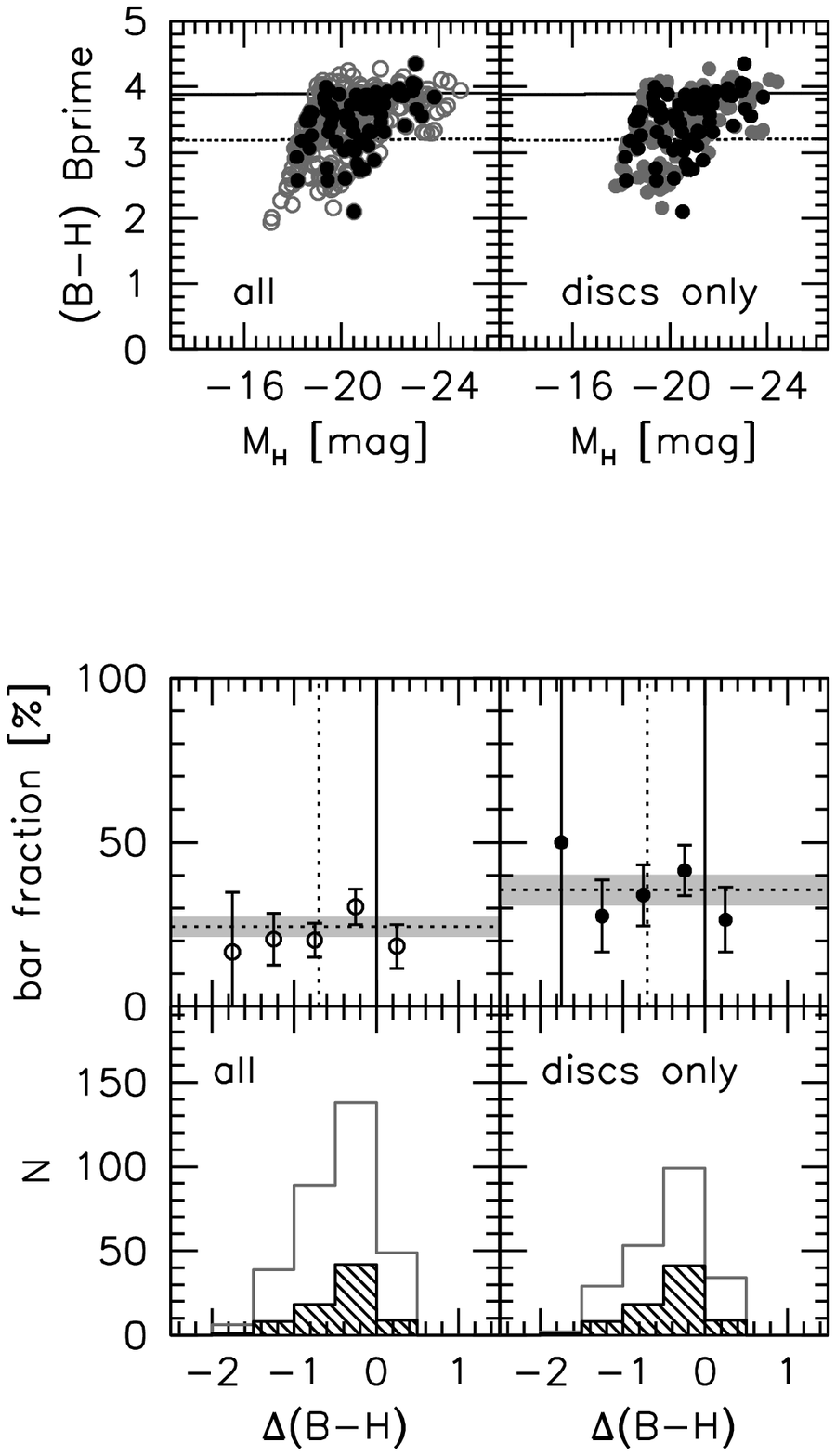}
\caption{{\it Top:} Color-magnitude (CM) distribution of the galaxies
  in the \Bprime\ sample for all morphological types (left) and for
  discs only (right); open symbols are all galaxies and filled symbols
  are the barred galaxies.  The solid line is the CM relation fit to
  the merged \scprime\ sample and the dotted line denotes
  $2\sigma_{MAD}$ (see \S\ref{sec:cmd} for details); we use the latter
  to separate red sequence and blue cloud galaxies. {\it Bottom:}
  Distribution of the color deviation $\Delta(B-H)$ from the CM
  relation for all members (left) and discs only (right); the open
  histograms are all galaxies and the shaded histograms the barred
  galaxies.  The sub-panels show the barred fraction as a function of
  $\Delta(B-H)$; the solid vertical line denotes no color deviation
  and the dotted vertical line denotes the separation between red
  sequence and blue cloud members ($2\sigma_{MAD}$). The barred
  galaxies have the same color distribution as the total \Bprime\
  population.}
\label{fig:11}
\end{figure} 

\begin{figure}    
\centering
\includegraphics[width=0.40\textwidth]{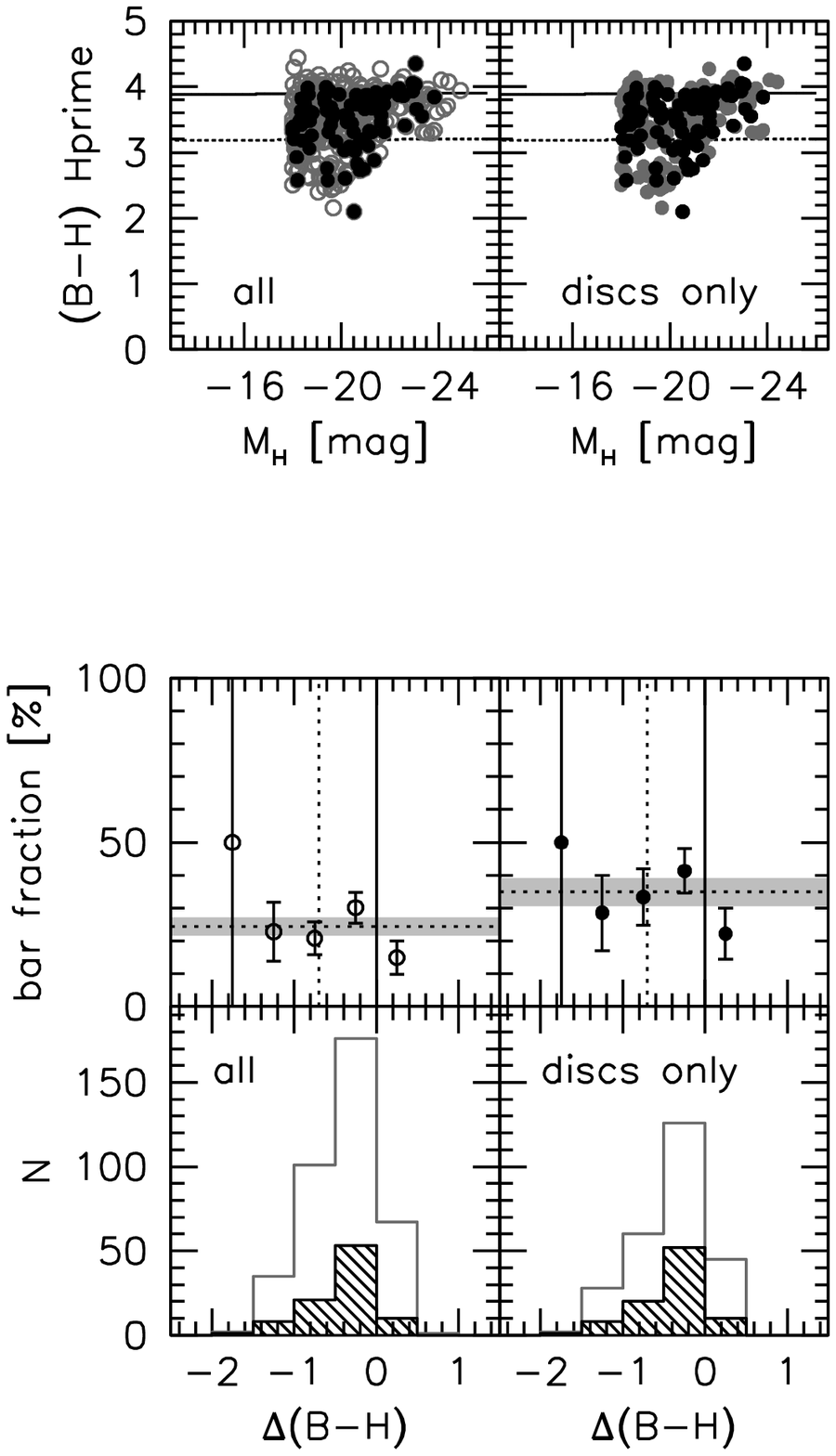}
\caption{Same as Figure \ref{fig:11} but for the \Hprime\ sample.
Again, we find that the barred galaxies have the same color
distribution as the total population.} 
\label{fig:12}
\end{figure} 

\begin{figure}    
\centering
\includegraphics[width=0.40\textwidth]{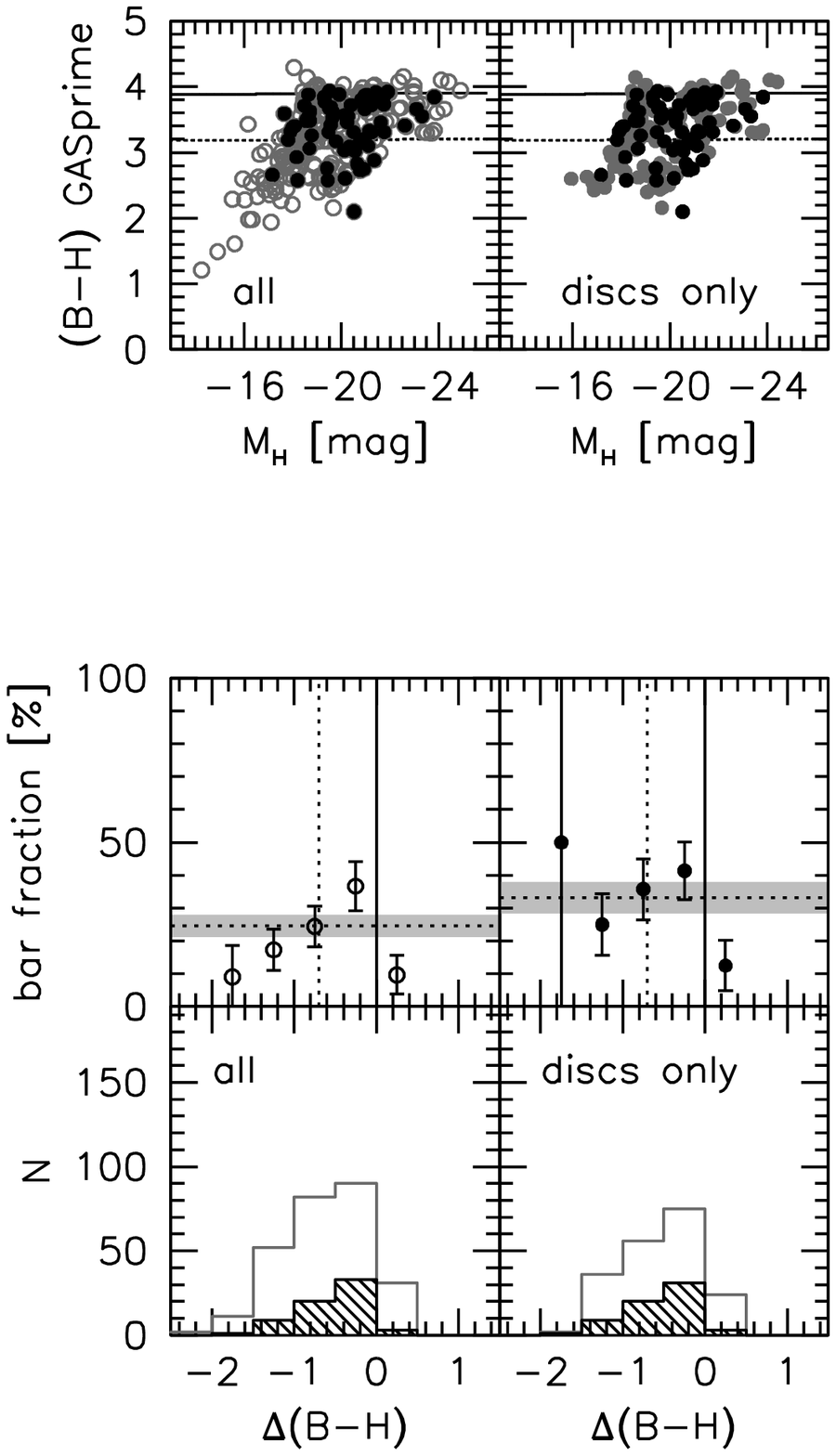}
\caption{Same as Figures \ref{fig:11} \& \ref{fig:12} but for the
\HIprime\ sample.  There are many more gas-rich irregular galaxies
that populate the faint blue region of the CM diagram.  Most of these
irregular galaxies are not barred which means that the barred fraction
decreases when considering the bluest galaxies in the \HIprime\ sample
(left panels).  However, this effect disappears when only discs
({\it S0--Sm}) are considered (right panels).}
\label{fig:13}
\end{figure}

\subsection{Color-Magnitude Diagram}
\label{sec:cmd}

To investigate the color distribution of barred versus non-barred
Virgo members, we first determine the color-magnitude (CM) relation
using $BH$ photometry.  We fit a least-squares to the CM relation
using a merged \scprime\ catalog and iteratively clip galaxies with a
color deviation $\Delta(B-H)$ larger than
($2\sigma_{MAD}$)\footnote{In case of heavy-tailed distributions, the
  most robust way of fitting the red-sequence is using the Median
  Absolute Deviation (MAD) as a scale indicator and using the MAD's
  relative $\sigma$ \citep{Beers90}}. We measure a CM relation of:

\begin{equation}
(B-H) = -0.002 M_H + 3.855
\end{equation}

\noindent with $\sigma_{MAD}\sim0.35$\footnote{Note that the GOLDMine
  photometry is not corrected for reddening across the Virgo field,
  thus the CM relation that we measured is only
  valid for internal comparison between our \scprime\ samples.}.
  Galaxies are divided into a red sequence and a blue cloud by
  measuring each galaxy's color deviation $\Delta(B-H)$ from the
  fitted CM relation and applying a color-cut of $2\sigma_{MAD}$.
  Figures~\ref{fig:11}, \ref{fig:12}, \& \ref{fig:13} show the
  color-magnitude diagrams for our three \scprime\ samples with the
  fitted CM relation as well as the $\Delta(B-H)$
  distributions for the total and barred galaxy populations.

If blue (disc-dominated) spirals have a higher barred fraction
(\citealt{Barazza08}), barred galaxies on average should then have
bluer colors. However, we find that the barred galaxies have the same
color distribution as the total galaxy population in both \Bprime\ and
\Hprime\ samples. This is true when considering all morphological
types (Figures~\ref{fig:11} \& \ref{fig:12}, left panels) as well as
only disc members (right panels). The barred fraction also stays
essentially constant as a function of $\Delta(B-H)$, i.e. the barred
fraction is not significantly higher in the blue cloud
(Table~\ref{tab:05}).

In contrast, the barred fraction in the \HIprime\ sample does depend
on color: when considering all morphological types
(Figure~\ref{fig:13}, left panels), the significant number of gas-rich
irregular galaxies ($\sim 1/3$ of the entire sample; see
\S\ref{sec:luminosity}, Figure~\ref{fig:07}, and
Table~\ref{tab:01}) populating the faint blue region of the CM
diagram results in a lower barred fraction for bluer galaxies. We find
that the irregular galaxies in the \HIprime\ sample all lie in the
blue cloud and that less than 1\% are barred including the irregulars
causes the global barred fraction to drop from $\sim29$\% on the red
sequence to $\sim19$\% in the blue cloud (Table~\ref{tab:05}).

Although both the \Bprime\ and \Hprime\ samples have dwarf galaxies
that also have a low barred fraction (see \S\ref{sec:luminosity}), the
dwarf galaxies span a range in color and include red sequence members
while the irregulars in the \HIprime\ sample are all blue. When only
disc members are considered, i.e. when dwarf and irregular galaxies
are excluded, the barred fraction in the \HIprime\ sample behaves as
in the \Bprime\ and \Hprime\ disc-only samples (Figure~\ref{fig:13},
right panels).

In summary, the barred fraction in the luminosity-selected samples
(\Bprime\ \& \Hprime) has the same color distribution as the total
population, and the barred fraction does not vary with color. The
barred fraction of $\sim21-26$\% (Table~\ref{tab:05}) on both
the red sequence and the blue cloud is the same as the value measured
for the total galaxy population ($\sim23$\%; Table~\ref{tab:02}),
i.e. using color as a proxy for morphology and measuring the barred
fraction in the blue cloud results in lower barred fraction than that
measured using morphologically classified discs ($\sim34$\%;
Table~\ref{tab:02}).  The only color dependence is in the
\HIprime\ sample where the sample's significant number of irregular
galaxies means that the barred fraction is lower at bluer colors,
i.e. in the blue cloud.

\subsection{Properties of Barred Lenticular \& Spiral Galaxies}
\label{sec:properties} 

\begin{figure}  
\centering
\includegraphics[width=0.45\textwidth]{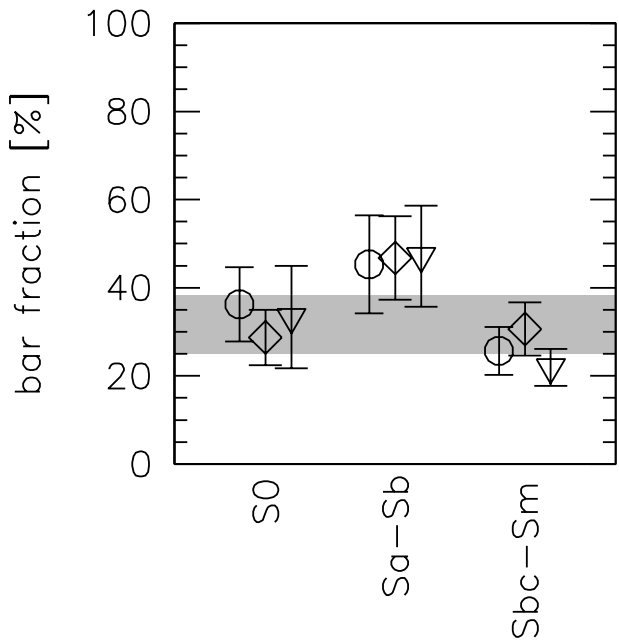}
\caption{Barred fraction for all disc galaxies divided into
  morphological subclasses: lenticulars ({\it S0--S0/a}), early-type
  spirals ({\it Sa--Sb}) and late-type spirals ({\it Sbc--Sm}).  The
  \Bprime\ sample is shown as circles, the \Hprime\ sample as
  diamonds, and the \HIprime\ sample as triangles; points are slightly
  offset horizontally for clarity, and statistical Poisson error-bars
  are included. The shaded horizontal region represents the range in
  barred fraction of the \scprime\ disc-only populations.  In all
  three \scprime\ samples, the early-type spirals have a measurably
  higher barred fraction relative to the lenticulars and late-type
  spirals; from statistical tests, we confirm that this result is
  significant at the $3\sigma$ level (see \S\ref{sec:properties}).}
\label{fig:14}
\end{figure}

Our analysis thus far highlights the importance of including
morphology (discs vs. non-discs) to identify trends in the barred
population. To further test for any trends in the disc population, we
divide our disc samples into lenticular ({\it S0--S0/a}), early-type
spirals ({\it Sa--Sb}), and late-type spirals ({\it Sbc--Sm}). We find
that the barred fraction is uniformly highest in the early-type
spirals (Table~\ref{tab:06} \& Figure~\ref{fig:14}): nearly
half of the early-type spirals (ETS) are barred compared to less than
one third of late-type spirals (LTS).

We test the robustness of this result by generating $10^4$ fake
datasets using the \Bprime\ disc sample, keeping the bar fraction
fixed as well as the ratio between the three different morphological
subclasses (i.e., lenticulars, early- and late-type spirals).  For
each realization, we randomly pair a 1) disc galaxy and 2) bar/non-bar
classification from the \Bprime\ disc sample; each realization
contains $N_D$ galaxies, where $N_D$ is the total number of \Bprime\
discs.  Only $\sim$0.3\% realizations have a barred early-type spiral
fraction of $(BF)_{ETS}\geq45.3$\% and a barred late-type spiral
fraction of $(BF)_{LTS}\leq25.7$\%, i.e. the observed difference in
the barred fractions between early- and late-type spirals
(Table~\ref{tab:06}) is significant at the $3\sigma$ level.

Repeating this exercise for the \Hprime\ and \HIprime\ disc samples
confirms that the early-type spirals have a higher barred fraction
compared to the late-type spirals ($3\sigma$ significance).  Our
statistical analysis supports the general conclusion that the barred
fraction is highest in early-type spirals in all three \scprime\
samples and that it is different from the barred fraction in late-type
spirals.

With the CM relation, we can also test whether the early-type spirals
differ in color from the lenticulars and late-type spirals; the
average $(B-H)$ color for each disc population is listed in
Table~\ref{tab:06}.  Not surprisingly, the lenticulars mostly
lie on the red sequence defined by the fitted CM relation.  However,
we find that the early-type spirals are nearly as red as the
lenticulars while the late-type spirals are significantly bluer; this
is true in all of the \scprime\ samples (Table~\ref{tab:06}).
In general, the early-type spirals in Virgo are more likely to be
barred and are redder than late-type spirals.  We note that
\citet{Masters09} also find that red spirals tend to be more barred
than blue spirals using Galaxy Zoo, a survey that spans a wide range
in environment at $z<0.085$ \citep{Raddick2007}.

Our results differ from \cite{Barazza08} who find that the barred
fraction increases from $\sim40$\% in galaxies with prominent bulges
to $\sim70$\% in disc-dominated galaxies.  However, the authors
separate bulge and disc-dominated galaxies parametrically by using,
e.g. color and/or S\'ersic index, and we have shown that the barred
fraction measured in a color-selected sample differs from that of a
morphologically defined sample (see \S\ref{sec:cmd}).  The differences
in barred fraction may also be due to using ellipse-fitting in
$r$-band images to identify bars \citep{Barazza08} versus visual
identification with H-band imaging (this work).  Note that the
\cite{Barazza08} sample contains mostly field galaxies while we focus
here on the richer environment of Virgo; we explore the environment
issue in Paper II of this series.

\begin{deluxetable*}{lcccccc}   
\tabletypesize{\scriptsize}
\tablewidth{0.80\textwidth}
\tablecaption{Disc Galaxies: Barred Fractions\tablenotemark{a} and Colors}
\tablehead{
\colhead{Sample} & \colhead{$N$} & \colhead{$N_{Bar}$} &
\colhead{Barred \%} & \colhead{$\mu_{1/2} (B-H)$\tablenotemark{b}}
}
\startdata
  \Bprime\  ({\it S0--Sm})                   & 231 & 77 & 33.3$\pm$ 4.4\% & 3.54\\
  \Bprime\  ({\it S0--S0/a})\tablenotemark{c} &  69 & 25 & 36.2$\pm$ 8.5\% & 3.79\\
  \Bprime\  ({\it Sa--Sb})\tablenotemark{d}  &  53 & 24 & 45.3$\pm$11.1\% & 3.71\\
  \Bprime\  ({\it Sbc--Sm})\tablenotemark{e} & 109 & 28 & 25.7$\pm$ 5.4\% & 3.16\\
  \hline
  \Hprime\  ({\it S0--Sm})                   & 282 & 97 & 34.4$\pm$ 4.0\% & 3.57\\
  \Hprime\  ({\it S0--S0/a})\tablenotemark{c} &  94 & 27 & 28.7$\pm$ 6.3\% & 3.80\\
  \Hprime\  ({\it Sa--Sb})\tablenotemark{d}  &  77 & 36 & 46.8$\pm$ 9.4\% & 3.71\\
  \Hprime\  ({\it Sbc--Sm})\tablenotemark{e} & 111 & 34 & 30.6$\pm$ 6.0\% & 3.18\\
  \hline   
  \HIprime\ ({\it S0--Sm})                   & 241 & 70 & 29.0$\pm$ 3.9\% & 3.41\\
  \HIprime\ ({\it S0--S0/a})\tablenotemark{c} &  33 & 11 & 33.3$\pm$11.6\% & 3.87\\
  \HIprime\ ({\it Sa--Sb})\tablenotemark{d}  &  53 & 25 & 47.2$\pm$11.4\% & 3.72\\
  \HIprime\ ({\it Sbc--Sm})\tablenotemark{e} & 155 & 34 & 21.9$\pm$ 4.2\% & 3.13\\
\enddata
  \tablenotetext{a}{~All bar classifications assigned using UKIDSS
    H-band imaging.}
\tablenotetext{b}{~The median $(B-H)$ color for these galaxies.}
\tablenotetext{c}{~Lenticulars.}
\tablenotetext{d}{~Early-type spirals.}
\tablenotetext{e}{~Late-type spirals.}
\label{tab:06}
\end{deluxetable*}

\section{Discussion}
\label{sec:discussion}

\subsection{Dependence on Galaxy Morphology}

Our study of barred galaxies in the Virgo cluster explores in detail
how the barred fraction varies over a wide range in luminosity, HI gas
mass mass, morphology, and color.  Using UKIDSS H-band imaging to
identify bars, we find that the barred fraction depends most strongly
on the morphological composition of the sample: for disc galaxies, the
barred fraction is $29-34$\% in both the luminosity-selected samples
(\Bprime, \Hprime) and the HI gas mass selected sample (\HIprime;
Table~\ref{tab:02}).  The disc barred fraction does not vary
strongly with luminosity nor HI gas mass (Table~\ref{tab:04};
Figures~\ref{fig:08}, \ref{fig:09}, \& \ref{fig:10}).  

However, the barred fraction in our \scprime\ samples drops to
$17-24$\% when we include all morphological types.  The lower barred
fractions are due to 1) the ellipticals that make up $\sim3-11$\% of
the \scprime\ samples and 2) the dwarf and irregular galaxies that
make up $8-34$\% of the \scprime\ samples (Table~\ref{tab:01}).  In
Virgo, we find that less than 2\% of the dwarf galaxies are barred,
and none of the irregular galaxies are barred.  Because our study
spans a wide range in luminosity and HI gas mass, a large number of
these low mass systems are included in our samples, and essentially
none of these systems have bars.

The impact of these low mass galaxies can be seen when comparing the
barred fraction versus luminosity or HI gas: there is a break in the
barred fraction at lower luminosities (Figures~\ref{fig:08} \&
\ref{fig:09}) and HI gas mass (Figure~\ref{fig:10}).  This effect
is particularly strong in the \HIprime\ sample because there are as
many irregular galaxies as late-type spirals (Figure~\ref{fig:07}).

The inclusion of low mass galaxies will also lower the barred fraction
in color-selected samples.  If we use color alone to select late-type
galaxies, the sample will naturally include a number of irregular
galaxies (e.g. Figure~\ref{fig:13}).  The measured barred fraction in
blue, presumably disc-dominated systems therefore will be lower than
the barred fraction for a morphologically selected disc sample.

A surprising result of our study is that we do not find as high a
barred fraction for disc galaxies as E00 ($\sim70$\%) even though both
studies use NIR imaging; the highest barred fraction
measured in our \scprime\ samples is for the early-type spirals
($\sim45-50$\%).  However, most of the 186 galaxies in the E00 sample
are luminous ($L>L^{\ast}$) discs in the field while our \scprime\
samples span a wide range in luminosity (factor of $\sim100$) and HI
gas mass, and we focus only on Virgo members.  A more extensive
analysis of the barred fraction as a function of environment is in
Paper II of this series.

\subsection{Barred Fractions of Disc Galaxies}

To better understand what physical mechanism sets the barred fraction,
we examine how the barred fraction varies in the disc population; note
that $>93$\% of the bars are in disc galaxies
(Table~\ref{tab:02}).  We find that in all three \scprime\
samples, the early-type spirals have the highest barred fraction
($\sim45-50$\%) while both the lenticulars and late-type spirals have
lower barred fractions ($22-36$\%; Table~\ref{tab:06} \&
Figure~\ref{fig:14}).  The barred early-type spirals are also
nearly as red as the barred lenticular galaxies, and both are
measurably redder than the barred late-type spirals
(Table~\ref{tab:06}), i.e. barred galaxies are not
preferentially bluer than non-barred galaxies.

The difference in barred fraction across the disc population is
surprising because in numerical simulations bars are 1) easily
triggered through gravitational interactions with other galaxies
\citep{Toth1992, Benson2004, Dubinski2008} and 2) difficult to destroy
even as the galaxy builds up its central mass \citep{Shen04,
  Athanassoula05, Debattista2006}.  Thus the barred fraction should not
depend on the bulge component.

However, the barred fraction may be higher in early-type spirals due
to their higher baryon fractions compared to late-type spirals.  The
susceptibility of galactic discs to global non-axisymmetric
instabilities is the $X$ parameter \citep{Goldreich1978, Toomre1981},
which for $m=2$ bar instabilities is:

\begin{equation}
X={{\kappa^2R}\over{4\pi G \Sigma}} 
\end{equation}

\noindent where $\kappa$ is the epicyclic frequency, $R$ is the radius
and the disc surface density is $\Sigma$.  The inverse dependence on
$\Sigma$ means that `maximum' or heavy discs such as those in
early-type spirals are more likely to undergo a bar instability
compared to late-type discs that tend to be less luminous, be in
smaller dark matter halos, and have lower baryon fractions
\citep{McGaugh2000}.

Environment is also likely to have a role, e.g. gravitational
interactions with other galaxies and/or the cluster potential can
trigger bar instabilities \citep{Moore1996Nature} as well as transform
late-type spirals into dwarf galaxies via galaxy harassment
\citep{Moore1999, Mastropietro05}.  In this scenario, an in-falling
late-type spiral undergoes a bar instability during the initial stage
of this transformation, but as its stars are stripped and the
remaining stellar component heated by gravitational encounters, the
galaxy is no longer a late-type spiral.  Late-type spirals would
become dwarfs in less than a cluster crossing time which, if true,
implies that most of the Sc/Sd galaxies in our Virgo cluster sample
have yet to travel deep into the cluster potential.

In contrast, the early-type spirals can survive a cluster crossing
because they can respond adiabatically to gravitational encounters and
are thus less likely to be transformed into dwarfs.  However, their
discs are heated and in the process become thicker and less
susceptible to bar instabilities \citep{Moore1999}, i.e. the
early-type spirals can evolve into lenticulars.  Note that the
early-type spirals are already as red as the lenticulars, thus their
luminosity-weighted ages would be comparable.

\section{Conclusions}
\label{sec:conclusions}

We study in detail how the barred galaxy fraction varies as a function
of luminosity, morphology, color, and HI gas mass in the Virgo cluster
by combining the Virgo Cluster Catalog \citep{Binggeli1985_P2} with
multiple public data-sets including recently released NIR
imaging from UKIDSS \citep{Lawrence2007}, HI gas masses from ALFALFA
\citep{Giovanelli2005_ALFALFA}, and photometry from GOLDMine
\citep{Gavazzi2003_GM}.  We define three \scprime\ samples where
galaxies are selected by their B-band luminosity (343; \Bprime),
H-band luminosity (413; \Hprime), and HI gas mass (439; \HIprime).
Bars are visually assigned using the high resolution H-band imaging
from UKIDSS; highly inclined systems are excluded.

For morphologically selected discs, the barred fraction in Virgo is
$\sim29-34$\% in all three of our \scprime\ samples, i.e. the barred
disc fraction does not depend strongly on how the sample is defined.
The barred disc fraction is surprisingly robust: it shows little
variation with luminosity or HI gas mass.  We also do not find any
evidence of barred galaxies being preferentially blue: the barred
galaxies have the same $(B-H)$ color distribution as the total
\scprime\ populations.

We find that the barred fraction depends most strongly on
morphological composition: when all galaxies in the \scprime\ samples
are included, the barred fraction drops to $\sim17-24$\%.  The lower
barred fraction relative to discs-only is due to ellipticals as well
as numerous dwarf and irregular galaxies that are included because of
our wide ranges in luminosity (factor $\sim100$) and HI gas mass
($M_{HI}>10^{7.5}M_{\odot}$).  Essentially none of the dwarf and
irregular galaxies are barred, but they make up to, e.g. 34\% of the
\HIprime\ sample.  These low-mass galaxies cause the barred fraction
to decrease at low luminosities/HI gas masses.  For studies that use
color alone to select blue, presumably disc-dominated galaxies, the
dwarf and irregular galaxies will lower the measured barred fraction
compared to that for a morphologically selected disc sample.

When the disc populations are separated into lenticulars ({\it S0--S0/a}),
early-type spirals ({\it Sa--Sb}), and late-type spirals ({\it Sbc--Sm}), we find
that the early-type spirals have a higher barred fraction
($\sim45-50$\%) compared to the lenticulars and late-type spirals
($\sim22-36$\%).  Statistical tests confirm that the difference in the
barred fractions is significant at the $3\sigma$ level.  The
early-type spirals are as red in $(B-H)$ color as the lenticulars, and
both are redder than the late-type spirals.

A possible explanation for the higher barred fraction in early-type
spirals and their red colors is that while bars are easily triggered
in disc galaxies, only discs with large baryon fractions/bulges
survive passage through the cluster potential.  The early-type spirals
form bars but lose their gas to the intra-cluster medium and so stop
forming new stars.  Gravitational interactions with other galaxies and
the cluster potential eventually heat the disc which then dissolves
the bar, and the early-type spiral evolves into a lenticular galaxy.
Bars also form in the late-type spirals, but these galaxies are
transformed by gravitational encounters into dwarf galaxies in less
than a cluster crossing time.

We note that the barred disc fraction in our \scprime\ samples is half
that measured by E00 using NIR imaging of local disc
galaxies.  However, the E00 sample is dominated by ($L>L^{\ast}$)
field galaxies while our study focuses on galaxies spanning a range in
luminosity in the richer environment of Virgo.  To determine if the
barred fraction varies with environment, we compare our Virgo results
to a field sample in Paper II of this series.

\begin{acknowledgments}
  We thank V. Debattista and S. Jogee for helpful discussion. This
  paper has made use of the GOLD Mine database and of HI measurements
  collected at the Arecibo Observatory for the ALFALFA survey. The
  Arecibo Observatory is part of the National Astronomy and Ionosphere
  Center which is operated by Cornell University under a cooperative
  agreement with the National Science Foundation. We are grateful for
  access to the ALFALFA data and acknowledge the leadership of the
  ALFALFA survey's Principal Investigators Martha Haynes and Riccardo
  Giovanelli, and the valuable contributions of the other ALFALFA team
  members.  L.G., K.T., and A.S. acknowledge generous support from the
  Swiss National Science Foundation (grant PP002-110576).
\end{acknowledgments}

\clearpage

\clearpage


\begin{thebibliography}{62}
\expandafter\ifx\csname natexlab\endcsname\relax\def\natexlab#1{#1}\fi

\bibitem[{{Aguerri} {et~al.}(2009){Aguerri}, {M{\'e}ndez-Abreu}, \&
  {Corsini}}]{Aguerri2009}
{Aguerri}, J.~A.~L., {M{\'e}ndez-Abreu}, J., \& {Corsini}, E.~M. 2009, \aap,
  495, 491

\bibitem[{{Athanassoula}(2005)}]{Athanassoula2005}
{Athanassoula}, E. 2005, \mnras, 358, 1477

\bibitem[{{Athanassoula} {et~al.}(2005){Athanassoula}, {Lambert}, \&
  {Dehnen}}]{Athanassoula05}
{Athanassoula}, E., {Lambert}, J.~C., \& {Dehnen}, W. 2005, \mnras, 363, 496

\bibitem[{{Balogh} {et~al.}(2001){Balogh}, {Christlein}, {Zabludoff}, \&
  {Zaritsky}}]{Balogh01}
{Balogh}, M.~L., {Christlein}, D., {Zabludoff}, A.~I., \& {Zaritsky}, D. 2001,
  \apj, 557, 117

\bibitem[{{Barazza} {et~al.}(2002){Barazza}, {Binggeli}, \&
  {Jerjen}}]{Barazza02}
{Barazza}, F.~D., {Binggeli}, B., \& {Jerjen}, H. 2002, \aap, 391, 823

\bibitem[{{Barazza} {et~al.}(2009){Barazza}, {Jablonka}, {Desai}, {Jogee},
  {Arag{\'o}n-Salamanca}, {De Lucia}, {Saglia}, {Halliday}, {Poggianti},
  {Dalcanton}, {Rudnick}, {Milvang-Jensen}, {Noll}, {Simard}, {Clowe},
  {Pell{\'o}}, {White}, \& {Zaritsky}}]{Barazza2009}
{Barazza}, F.~D., {Jablonka}, P., {Desai}, V., {Jogee}, S.,
  {Arag{\'o}n-Salamanca}, A., {De Lucia}, G., {Saglia}, R.~P., {Halliday}, C.,
  {Poggianti}, B.~M., {Dalcanton}, J.~J., {Rudnick}, G., {Milvang-Jensen}, B.,
  {Noll}, S., {Simard}, L., {Clowe}, D.~I., {Pell{\'o}}, R., {White}, S.~D.~M.,
  \& {Zaritsky}, D. 2009, \aap, 497, 713

\bibitem[{{Barazza} {et~al.}(2008){Barazza}, {Jogee}, \&
  {Marinova}}]{Barazza08}
{Barazza}, F.~D., {Jogee}, S., \& {Marinova}, I. 2008, \apj, 675, 1194

\bibitem[{{Beers} {et~al.}(1990){Beers}, {Flynn}, \& {Gebhardt}}]{Beers90}
{Beers}, T.~C., {Flynn}, K., \& {Gebhardt}, K. 1990, \aj, 100, 32

\bibitem[{{Bell} \& {de Jong}(2001)}]{Bell01}
{Bell}, E.~F., \& {de Jong}, R.~S. 2001, \apj, 550, 212

\bibitem[{{Benson} {et~al.}(2004){Benson}, {Lacey}, {Frenk}, {Baugh}, \&
  {Cole}}]{Benson2004}
{Benson}, A.~J., {Lacey}, C.~G., {Frenk}, C.~S., {Baugh}, C.~M., \& {Cole}, S.
  2004, \mnras, 351, 1215

\bibitem[{{Bertin} {et~al.}(2002){Bertin}, {Mellier}, {Radovich}, {Missonnier},
  {Didelon}, \& {Morin}}]{Bertin2002}
{Bertin}, E., {Mellier}, Y., {Radovich}, M., {Missonnier}, G., {Didelon}, P.,
  \& {Morin}, B. 2002, in Astronomical Society of the Pacific Conference
  Series, Vol. 281, Astronomical Data Analysis Software and Systems XI, ed.
  {D.~A.~Bohlender, D.~Durand, \& T.~H.~Handley}, 228--+

\bibitem[{{Binggeli}(1999)}]{Binggeli99}
{Binggeli}, B. 1999, in Lecture Notes in Physics, Berlin Springer Verlag, Vol.
  530, The Radio Galaxy Messier 87, ed. {H.-J.~R{\"o}ser \& K.~Meisenheimer},
  9--+

\bibitem[{{Binggeli} {et~al.}(1985){Binggeli}, {Sandage}, \&
  {Tammann}}]{Binggeli1985_P2}
{Binggeli}, B., {Sandage}, A., \& {Tammann}, G.~A. 1985, \aj, 90, 1681

\bibitem[{{Binggeli} {et~al.}(1984){Binggeli}, {Sandage}, \&
  {Tarenghi}}]{Binggeli1984_P1}
{Binggeli}, B., {Sandage}, A., \& {Tarenghi}, M. 1984, \aj, 89, 64

\bibitem[{{B{\"o}ker} {et~al.}(2003){B{\"o}ker}, {Stanek}, \& {van der
  Marel}}]{Boeker2003}
{B{\"o}ker}, T., {Stanek}, R., \& {van der Marel}, R.~P. 2003, \aj, 125, 1073

\bibitem[{{Bureau} \& {Athanassoula}(2005)}]{Bureau2005}
{Bureau}, M., \& {Athanassoula}, E. 2005, \apj, 626, 159

\bibitem[{{Carollo} {et~al.}(1997){Carollo}, {Stiavelli}, {de Zeeuw}, \&
  {Mack}}]{Carollo1997_P1}
{Carollo}, C.~M., {Stiavelli}, M., {de Zeeuw}, P.~T., \& {Mack}, J. 1997, \aj,
  114, 2366

\bibitem[{{Carollo} {et~al.}(1998){Carollo}, {Stiavelli}, \&
  {Mack}}]{Carollo1998_P2}
{Carollo}, C.~M., {Stiavelli}, M., \& {Mack}, J. 1998, \aj, 116, 68

\bibitem[{{Casali} {et~al.}(2007){Casali}, {Adamson}, {Alves de Oliveira},
  {Almaini}, {Burch}, {Chuter}, {Elliot}, {Folger}, {Foucaud}, {Hambly},
  {Hastie}, {Henry}, {Hirst}, {Irwin}, {Ives}, {Lawrence}, {Laidlaw}, {Lee},
  {Lewis}, {Lunney}, {McLay}, {Montgomery}, {Pickup}, {Read}, {Rees}, {Robson},
  {Sekiguchi}, {Vick}, {Warren}, \& {Woodward}}]{Casali2007}
{Casali}, M., {Adamson}, A., {Alves de Oliveira}, C., {Almaini}, O., {Burch},
  K., {Chuter}, T., {Elliot}, J., {Folger}, M., {Foucaud}, S., {Hambly}, N.,
  {Hastie}, M., {Henry}, D., {Hirst}, P., {Irwin}, M., {Ives}, D., {Lawrence},
  A., {Laidlaw}, K., {Lee}, D., {Lewis}, J., {Lunney}, D., {McLay}, S.,
  {Montgomery}, D., {Pickup}, A., {Read}, M., {Rees}, N., {Robson}, I.,
  {Sekiguchi}, K., {Vick}, A., {Warren}, S., \& {Woodward}, B. 2007, \aap, 467,
  777

\bibitem[{{Combes} {et~al.}(1990){Combes}, {Debbasch}, {Friedli}, \&
  {Pfenniger}}]{Combes1990}
{Combes}, F., {Debbasch}, F., {Friedli}, D., \& {Pfenniger}, D. 1990, \aap,
  233, 82

\bibitem[{{Courteau} {et~al.}(2003){Courteau}, {Andersen}, {Bershady},
  {MacArthur}, \& {Rix}}]{Courteau2003}
{Courteau}, S., {Andersen}, D.~R., {Bershady}, M.~A., {MacArthur}, L.~A., \&
  {Rix}, H.-W. 2003, \apj, 594, 208

\bibitem[{{de Vaucouleurs}(1977)}]{deVaucouleurs1977}
{de Vaucouleurs}, G. 1977, \nat, 266, 126

\bibitem[{{de Vaucouleurs} {et~al.}(1991){de Vaucouleurs}, {de Vaucouleurs},
  {Corwin}, {Buta}, {Paturel}, \& {Fouque}}]{RC3}
{de Vaucouleurs}, G., {de Vaucouleurs}, A., {Corwin}, Jr., H.~G., {Buta},
  R.~J., {Paturel}, G., \& {Fouque}, P. 1991, {Third Reference Catalogue of
  Bright Galaxies} (de Vaucouleurs, G., de Vaucouleurs, A., Corwin, H. G., Jr.,
  Buta, R. J., Paturel, G., \& Fouque, P.)

\bibitem[{{Debattista} {et~al.}(2006){Debattista}, {Mayer}, {Carollo}, {Moore},
  {Wadsley}, \& {Quinn}}]{Debattista2006}
{Debattista}, V.~P., {Mayer}, L., {Carollo}, C.~M., {Moore}, B., {Wadsley}, J.,
  \& {Quinn}, T. 2006, \apj, 645, 209

\bibitem[{{Dubinski} {et~al.}(2008){Dubinski}, {Gauthier}, {Widrow}, \&
  {Nickerson}}]{Dubinski2008}
{Dubinski}, J., {Gauthier}, J., {Widrow}, L., \& {Nickerson}, S. 2008, in
  Astronomical Society of the Pacific Conference Series, Vol. 396, Astronomical
  Society of the Pacific Conference Series, ed. {J.~G.~Funes \& E.~M.~Corsini},
  321--+

\bibitem[{{Eskridge} {et~al.}(2002){Eskridge}, {Frogel}, {Pogge}, {Quillen},
  {Berlind}, {Davies}, {Depoy}, {Gilbert}, {Houdashelt}, {Kuchinski},
  {Ramirez}, {Sellgren}, {Stutz}, {Terndrup}, \& {Tiede}}]{Eskridge2002}
{Eskridge}, P.~B., {Frogel}, J.~A., {Pogge}, R.~W., {Quillen}, A.~C.,
  {Berlind}, A.~A., {Davies}, R.~L., {Depoy}, D.~L., {Gilbert}, K.~M.,
  {Houdashelt}, M.~L., {Kuchinski}, L.~E., {Ramirez}, S.~V., {Sellgren}, K.,
  {Stutz}, A., {Terndrup}, D.~M., \& {Tiede}, G.~P. 2002, VizieR Online Data
  Catalog, 214, 30073

\bibitem[{{Eskridge} {et~al.}(2000){Eskridge}, {Frogel}, {Pogge}, {Quillen},
  {Davies}, {DePoy}, {Houdashelt}, {Kuchinski}, {Ram{\'{\i}}rez}, {Sellgren},
  {Terndrup}, \& {Tiede}}]{Eskridge2000}
{Eskridge}, P.~B., {Frogel}, J.~A., {Pogge}, R.~W., {Quillen}, A.~C., {Davies},
  R.~L., {DePoy}, D.~L., {Houdashelt}, M.~L., {Kuchinski}, L.~E.,
  {Ram{\'{\i}}rez}, S.~V., {Sellgren}, K., {Terndrup}, D.~M., \& {Tiede}, G.~P.
  2000, \aj, 119, 536

\bibitem[{{Gavazzi} \& {Boselli}(1996)}]{Gavazzi1996}
{Gavazzi}, G., \& {Boselli}, A. 1996, Astrophysical Letters Communications, 35,
  1

\bibitem[{{Gavazzi} {et~al.}(2003){Gavazzi}, {Boselli}, {Donati}, {Franzetti},
  \& {Scodeggio}}]{Gavazzi2003_GM}
{Gavazzi}, G., {Boselli}, A., {Donati}, A., {Franzetti}, P., \& {Scodeggio}, M.
  2003, \aap, 400, 451

\bibitem[{{Gavazzi} {et~al.}(1999){Gavazzi}, {Boselli}, {Scodeggio}, {Pierini},
  \& {Belsole}}]{Gavazzi1999}
{Gavazzi}, G., {Boselli}, A., {Scodeggio}, M., {Pierini}, D., \& {Belsole}, E.
  1999, \mnras, 304, 595

\bibitem[{{Gavazzi} {et~al.}(2005){Gavazzi}, {Boselli}, {van Driel}, \&
  {O'Neil}}]{Gavazzi2005}
{Gavazzi}, G., {Boselli}, A., {van Driel}, W., \& {O'Neil}, K. 2005, \aap, 429,
  439

\bibitem[{{Gavazzi} {et~al.}(2008){Gavazzi}, {Giovanelli}, {Haynes}, {Fabello},
  {Fumagalli}, {Kent}, {Koopmann}, {Brosch}, {Hoffman}, {Salzer}, \&
  {Boselli}}]{Gavazzi2008}
{Gavazzi}, G., {Giovanelli}, R., {Haynes}, M.~P., {Fabello}, S., {Fumagalli},
  M., {Kent}, B.~R., {Koopmann}, R.~A., {Brosch}, N., {Hoffman}, G.~L.,
  {Salzer}, J.~J., \& {Boselli}, A. 2008, \aap, 482, 43

\bibitem[{{Gehrels}(1986)}]{Gehrels86}
{Gehrels}, N. 1986, \apj, 303, 336

\bibitem[{{Giovanelli} {et~al.}(2005){Giovanelli}, {Haynes}, {Kent},
  {Perillat}, {Saintonge}, {Brosch}, {Catinella}, {Hoffman}, {Stierwalt},
  {Spekkens}, {Lerner}, {Masters}, {Momjian}, {Rosenberg}, {Springob},
  {Boselli}, {Charmandaris}, {Darling}, {Davies}, {Garcia Lambas}, {Gavazzi},
  {Giovanardi}, {Hardy}, {Hunt}, {Iovino}, {Karachentsev}, {Karachentseva},
  {Koopmann}, {Marinoni}, {Minchin}, {Muller}, {Putman}, {Pantoja}, {Salzer},
  {Scodeggio}, {Skillman}, {Solanes}, {Valotto}, {van Driel}, \& {van
  Zee}}]{Giovanelli2005_ALFALFA}
{Giovanelli}, R., {Haynes}, M.~P., {Kent}, B.~R., {Perillat}, P., {Saintonge},
  A., {Brosch}, N., {Catinella}, B., {Hoffman}, G.~L., {Stierwalt}, S.,
  {Spekkens}, K., {Lerner}, M.~S., {Masters}, K.~L., {Momjian}, E.,
  {Rosenberg}, J.~L., {Springob}, C.~M., {Boselli}, A., {Charmandaris}, V.,
  {Darling}, J.~K., {Davies}, J., {Garcia Lambas}, D., {Gavazzi}, G.,
  {Giovanardi}, C., {Hardy}, E., {Hunt}, L.~K., {Iovino}, A., {Karachentsev},
  I.~D., {Karachentseva}, V.~E., {Koopmann}, R.~A., {Marinoni}, C., {Minchin},
  R., {Muller}, E., {Putman}, M., {Pantoja}, C., {Salzer}, J.~J., {Scodeggio},
  M., {Skillman}, E., {Solanes}, J.~M., {Valotto}, C., {van Driel}, W., \& {van
  Zee}, L. 2005, \aj, 130, 2598

\bibitem[{{Giovanelli} {et~al.}(2007){Giovanelli}, {Haynes}, {Kent},
  {Saintonge}, {Stierwalt}, {Altaf}, {Balonek}, {Brosch}, {Brown}, {Catinella},
  {Furniss}, {Goldstein}, {Hoffman}, {Koopmann}, {Kornreich}, {Mahmood},
  {Martin}, {Masters}, {Mitschang}, {Momjian}, {Nair}, {Rosenberg}, \&
  {Walsh}}]{Giovanelli2007}
{Giovanelli}, R., {Haynes}, M.~P., {Kent}, B.~R., {Saintonge}, A., {Stierwalt},
  S., {Altaf}, A., {Balonek}, T., {Brosch}, N., {Brown}, S., {Catinella}, B.,
  {Furniss}, A., {Goldstein}, J., {Hoffman}, G.~L., {Koopmann}, R.~A.,
  {Kornreich}, D.~A., {Mahmood}, B., {Martin}, A.~M., {Masters}, K.~L.,
  {Mitschang}, A., {Momjian}, E., {Nair}, P.~H., {Rosenberg}, J.~L., \&
  {Walsh}, B. 2007, \aj, 133, 2569

\bibitem[{{Goldreich} \& {Tremaine}(1978)}]{Goldreich1978}
{Goldreich}, P., \& {Tremaine}, S. 1978, \apj, 222, 850

\bibitem[{{Hambly} {et~al.}(2008){Hambly}, {Collins}, {Cross}, {Mann}, {Read},
  {Sutorius}, {Bond}, {Bryant}, {Emerson}, {Lawrence}, {Rimoldini}, {Stewart},
  {Williams}, {Adamson}, {Hirst}, {Dye}, \& {Warren}}]{Hambly2008}
{Hambly}, N.~C., {Collins}, R.~S., {Cross}, N.~J.~G., {Mann}, R.~G., {Read},
  M.~A., {Sutorius}, E.~T.~W., {Bond}, I., {Bryant}, J., {Emerson}, J.~P.,
  {Lawrence}, A., {Rimoldini}, L., {Stewart}, J.~M., {Williams}, P.~M.,
  {Adamson}, A., {Hirst}, P., {Dye}, S., \& {Warren}, S.~J. 2008, \mnras, 384,
  637

\bibitem[{{Hewett} {et~al.}(2006){Hewett}, {Warren}, {Leggett}, \&
  {Hodgkin}}]{Hewett2006}
{Hewett}, P.~C., {Warren}, S.~J., {Leggett}, S.~K., \& {Hodgkin}, S.~T. 2006,
  \mnras, 367, 454

\bibitem[{{Kalloghlian} \& {Kandalian}(1998)}]{Kalloghlian1998}
{Kalloghlian}, A.~T., \& {Kandalian}, R.~A. 1998, Astrophysics, 41, 119

\bibitem[{{Kent} {et~al.}(2009){Kent}, {Spekkens}, {Giovanelli}, {Haynes},
  {Momjian}, {Cort{\'e}s}, {Hardy}, \& {West}}]{Kent2009}
{Kent}, B.~R., {Spekkens}, K., {Giovanelli}, R., {Haynes}, M.~P., {Momjian},
  E., {Cort{\'e}s}, J.~R., {Hardy}, E., \& {West}, A.~A. 2009, \apj, 691, 1595

\bibitem[{{Laine} {et~al.}(1999){Laine}, {Kenney}, {Yun}, \&
  {Gottesman}}]{Laine1999}
{Laine}, S., {Kenney}, J.~D.~P., {Yun}, M.~S., \& {Gottesman}, S.~T. 1999,
  \apj, 511, 709

\bibitem[{{Laine} {et~al.}(2002){Laine}, {Shlosman}, {Knapen}, \&
  {Peletier}}]{Laine2002}
{Laine}, S., {Shlosman}, I., {Knapen}, J.~H., \& {Peletier}, R.~F. 2002, \apj,
  567, 97

\bibitem[{{Lawrence} {et~al.}(2007){Lawrence}, {Warren}, {Almaini}, {Edge},
  {Hambly}, {Jameson}, {Lucas}, {Casali}, {Adamson}, {Dye}, {Emerson},
  {Foucaud}, {Hewett}, {Hirst}, {Hodgkin}, {Irwin}, {Lodieu}, {McMahon},
  {Simpson}, {Smail}, {Mortlock}, \& {Folger}}]{Lawrence2007}
{Lawrence}, A., {Warren}, S.~J., {Almaini}, O., {Edge}, A.~C., {Hambly}, N.~C.,
  {Jameson}, R.~F., {Lucas}, P., {Casali}, M., {Adamson}, A., {Dye}, S.,
  {Emerson}, J.~P., {Foucaud}, S., {Hewett}, P., {Hirst}, P., {Hodgkin}, S.~T.,
  {Irwin}, M.~J., {Lodieu}, N., {McMahon}, R.~G., {Simpson}, C., {Smail}, I.,
  {Mortlock}, D., \& {Folger}, M. 2007, \mnras, 379, 1599

\bibitem[{{Lisker} {et~al.}(2007){Lisker}, {Grebel}, {Binggeli}, \&
  {Glatt}}]{Lisker07}
{Lisker}, T., {Grebel}, E.~K., {Binggeli}, B., \& {Glatt}, K. 2007, \apj, 660,
  1186

\bibitem[{{Marinova} \& {Jogee}(2007)}]{Marinova2007}
{Marinova}, I., \& {Jogee}, S. 2007, \apj, 659, 1176

\bibitem[{{Marinova} {et~al.}(2009){Marinova}, {Jogee}, {Heiderman}, {Barazza},
  {Gray}, {Barden}, {Wolf}, {Peng}, {Bacon}, {Balogh}, {Bell}, {B{\"o}hm},
  {Caldwell}, {H{\"a}u{\ss}ler}, {Heymans}, {Jahnke}, {van Kampen}, {Lane},
  {McIntosh}, {Meisenheimer}, {S{\'a}nchez}, {Somerville}, {Taylor},
  {Wisotzki}, \& {Zheng}}]{Marinova2009}
{Marinova}, I., {Jogee}, S., {Heiderman}, A., {Barazza}, F.~D., {Gray}, M.~E.,
  {Barden}, M., {Wolf}, C., {Peng}, C.~Y., {Bacon}, D., {Balogh}, M., {Bell},
  E.~F., {B{\"o}hm}, A., {Caldwell}, J.~A.~R., {H{\"a}u{\ss}ler}, B.,
  {Heymans}, C., {Jahnke}, K., {van Kampen}, E., {Lane}, K., {McIntosh}, D.~H.,
  {Meisenheimer}, K., {S{\'a}nchez}, S.~F., {Somerville}, R., {Taylor}, A.,
  {Wisotzki}, L., \& {Zheng}, X. 2009, \apj, 698, 1639

\bibitem[{{Martinez-Valpuesta} {et~al.}(2006){Martinez-Valpuesta}, {Shlosman},
  \& {Heller}}]{Martinez-Valpuesta2006}
{Martinez-Valpuesta}, I., {Shlosman}, I., \& {Heller}, C. 2006, \apj, 637, 214

\bibitem[{{Martini} \& {Pogge}(1999)}]{Martini1999}
{Martini}, P., \& {Pogge}, R.~W. 1999, \aj, 118, 2646

\bibitem[{{Masters} {et~al.}(2009){Masters}, {Mosleh}, {Romer}, {Nichol},
  {Bamford}, {Schawinski}, {Lintott}, {Andreescu}, {Campbell}, {Crowcroft},
  {Doyle}, {Edmondson}, {Murray}, {Raddick}, {Slosar}, {Szalay}, \&
  {Vandenberg}}]{Masters09}
{Masters}, K.~L., {Mosleh}, M., {Romer}, A.~K., {Nichol}, R.~C., {Bamford},
  S.~P., {Schawinski}, K., {Lintott}, C.~J., {Andreescu}, D., {Campbell},
  H.~C., {Crowcroft}, B., {Doyle}, I., {Edmondson}, E.~M., {Murray}, P.,
  {Raddick}, M.~J., {Slosar}, A., {Szalay}, A.~S., \& {Vandenberg}, J. 2009,
  ArXiv e-prints

\bibitem[{{Mastropietro} {et~al.}(2005){Mastropietro}, {Moore}, {Mayer},
  {Debattista}, {Piffaretti}, \& {Stadel}}]{Mastropietro05}
{Mastropietro}, C., {Moore}, B., {Mayer}, L., {Debattista}, V.~P.,
  {Piffaretti}, R., \& {Stadel}, J. 2005, \mnras, 364, 607

\bibitem[{{McGaugh} {et~al.}(2000){McGaugh}, {Schombert}, {Bothun}, \& {de
  Blok}}]{McGaugh2000}
{McGaugh}, S.~S., {Schombert}, J.~M., {Bothun}, G.~D., \& {de Blok}, W.~J.~G.
  2000, \apjl, 533, L99

\bibitem[{{Men{\'e}ndez-Delmestre} {et~al.}(2007){Men{\'e}ndez-Delmestre},
  {Sheth}, {Schinnerer}, {Jarrett}, \& {Scoville}}]{MenendezDelmestre2007}
{Men{\'e}ndez-Delmestre}, K., {Sheth}, K., {Schinnerer}, E., {Jarrett}, T.~H.,
  \& {Scoville}, N.~Z. 2007, \apj, 657, 790

\bibitem[{{Moore} {et~al.}(1996){Moore}, {Katz}, {Lake}, {Dressler}, \&
  {Oemler}}]{Moore1996Nature}
{Moore}, B., {Katz}, N., {Lake}, G., {Dressler}, A., \& {Oemler}, A. 1996,
  \nat, 379, 613

\bibitem[{{Moore} {et~al.}(1999){Moore}, {Lake}, {Quinn}, \&
  {Stadel}}]{Moore1999}
{Moore}, B., {Lake}, G., {Quinn}, T., \& {Stadel}, J. 1999, \mnras, 304, 465

\bibitem[{{Pfenniger} \& {Norman}(1990)}]{Pfenniger1990}
{Pfenniger}, D., \& {Norman}, C. 1990, \apj, 363, 391

\bibitem[{{Raddick} {et~al.}(2007){Raddick}, {Lintott}, {Schawinski}, {Thomas},
  {Nichol}, {Andreescu}, {Bamford}, {Land}, {Murray}, {Slosar}, {Szalay},
  {Vandenberg}, \& {Galaxy Zoo team}}]{Raddick2007}
{Raddick}, J., {Lintott}, C.~J., {Schawinski}, K., {Thomas}, D., {Nichol},
  R.~C., {Andreescu}, D., {Bamford}, S., {Land}, K.~R., {Murray}, P., {Slosar},
  A., {Szalay}, A.~S., {Vandenberg}, J., \& {Galaxy Zoo team}. 2007, in
  Bulletin of the American Astronomical Society, Vol.~38, Bulletin of the
  American Astronomical Society, 892--+

\bibitem[{{Roberts} \& {Haynes}(1994)}]{Roberts1994}
{Roberts}, M.~S., \& {Haynes}, M.~P. 1994, \araa, 32, 115

\bibitem[{{Shen} \& {Sellwood}(2004)}]{Shen04}
{Shen}, J., \& {Sellwood}, J.~A. 2004, \apj, 604, 614

\bibitem[{{Skrutskie} {et~al.}(2006){Skrutskie}, {Cutri}, {Stiening},
  {Weinberg}, {Schneider}, {Carpenter}, {Beichman}, {Capps}, {Chester},
  {Elias}, {Huchra}, {Liebert}, {Lonsdale}, {Monet}, {Price}, {Seitzer},
  {Jarrett}, {Kirkpatrick}, {Gizis}, {Howard}, {Evans}, {Fowler}, {Fullmer},
  {Hurt}, {Light}, {Kopan}, {Marsh}, {McCallon}, {Tam}, {Van Dyk}, \&
  {Wheelock}}]{2MASS}
{Skrutskie}, M.~F., {Cutri}, R.~M., {Stiening}, R., {Weinberg}, M.~D.,
  {Schneider}, S., {Carpenter}, J.~M., {Beichman}, C., {Capps}, R., {Chester},
  T., {Elias}, J., {Huchra}, J., {Liebert}, J., {Lonsdale}, C., {Monet}, D.~G.,
  {Price}, S., {Seitzer}, P., {Jarrett}, T., {Kirkpatrick}, J.~D., {Gizis},
  J.~E., {Howard}, E., {Evans}, T., {Fowler}, J., {Fullmer}, L., {Hurt}, R.,
  {Light}, R., {Kopan}, E.~L., {Marsh}, K.~A., {McCallon}, H.~L., {Tam}, R.,
  {Van Dyk}, S., \& {Wheelock}, S. 2006, \aj, 131, 1163

\bibitem[{{Toomre}(1981)}]{Toomre1981}
{Toomre}, A. 1981, in Structure and Evolution of Normal Galaxies, ed.
  {S.~M.~Fall \& D.~Lynden-Bell}, 111--136

\bibitem[{{Toth} \& {Ostriker}(1992)}]{Toth1992}
{Toth}, G., \& {Ostriker}, J.~P. 1992, \apj, 389, 5

\bibitem[{{van den Bergh}(2002)}]{VanDenBergh2002}
{van den Bergh}, S. 2002, \aj, 124, 782

\end{thebibliography}
\end{document}